\documentclass[english,a4paper,10pt,final]{article}

\usepackage[T1]{fontenc}
\usepackage[utf8]{inputenc}
\usepackage{array}
\usepackage{float}
\usepackage{amsmath}
\usepackage{amssymb}
\usepackage{graphicx}
\usepackage{bm}
\usepackage{subfigure}
\usepackage{algorithm}
\usepackage{babel}
\usepackage{authblk}

\makeatletter

\numberwithin{equation}{section}
\numberwithin{figure}{section}

\newtheorem{remark}{Remark}

\date{}
\makeatother

\begin{document}
\title{\bf Proportional-Integral Clock Synchronization \\ in Wireless Sensor Networks}

\author[1]{Kasım~Sinan~Yıldırım}
\author[2]{Ruggero~Carli}
\author[2]{Luca~Schenato}
\affil[1]{Department of Computer Engineering, Ege University, 35100, Bornova, İzmir, TURKEY, sinan.yildirim@ege.edu.tr}
\affil[2]{Department of Information Engineering, University of Padova, ITALY, \{carlirug,schenato\}@dei.unipd.it}

\maketitle

\begin{abstract}
In this article, we present a new control theoretic distributed time synchronization algorithm, named PISync, in order to synchronize sensor nodes in Wireless Sensor Networks (WSNs). PISync algorithm is based on a Proportional-Integral (PI) controller. It applies a proportional feedback (P) and an integral feedback (I) on the local measured synchronization errors to compensate the differences between the clock offsets and clock speeds. We present practical flooding-based and fully distributed protocol implementations of the PISync algorithm, and we provide theoretical analysis to highlight the benefits of this approach in terms of improved steady state error and scalability as compared to existing synchronization algorithms. We show through real-world experiments and simulations that PISync protocols have several advantages over existing protocols in the WSN literature, namely no need for memory allocation, minimal CPU overhead and code size independent of network size and topology, and graceful performance degradation in terms of network size.
\end{abstract}

\maketitle

\section{Introduction}

Built-in clocks of the sensor nodes are not sufficient alone to provide synchronized time notion in Wireless Sensor Networks (WSNs). This phenomenon arises from the fact that their crystal oscillators operate at different and unstable frequencies. On the other hand, being able to access to synchronized time notion is a prerequisite for the correct and efficient operation of WSNs \cite{Akyildiz2010}. In order to fulfil this requirement, sensor nodes are required to participate in \emph{time synchronization} periodically. At each synchronization round, they communicate with the other nodes to exchange time information and they process the received data to establish a function which represents the network-wide global time. However, various error sources arising from communication prevent from receiving time information perfectly. Moreover, limited resources and capabilities of the sensor nodes restrict the frequency of the communication as well as the computational and memory overhead of the information processing. These characteristics make time synchronization still an interesting and challenging problem in WSNs. Indeed, 
In the next section we provide a literature review of clock-synchronization in WSNs and the current state-of-the-art.  

\subsection{Literature review}

There is an ample body of studies in time synchronization in WSNs from both theoretical and practical perspectives. An initial gross classification of these works can be based whether the network synchronization is based on one or more reference nodes, referred as \emph{flooding-based} methods, or based on no reference node at all, referred as  \emph{fully-distributed} methods.

In the flooding-based methods, there is typically one node that is selected as reference for all other nodes either a-priori, as for example when stable time sources such as Coordinated Universal Time (UTC) time are available, or on-the-fly via leader election procedures. The reference node periodically disseminates its stable information through the network via \emph{flooding}, which is a common method to achieve network-wide time synchronization in WSNs thanks to its being a basic and simple communication primitive. Since there is no demand to construct and maintain a topological infrastructure, e.g. a spanning tree, flooding is also robust to dynamic topological changes. Each receiver node calculates its offset and frequency difference with respect to the received time information, and synchronizes its clock. There are several successful implementations of time synchronization with this simple and robust approach, as presented in \cite{Maroti2004,RATS,Lenzen2009Optimal,Schmid:2010,Yildirim2013,Yildirim2013_2}. These implementations calculate the offset and frequency difference of the clocks by employing \emph{least-squares regression} \cite{Wu:10}. However, this method has considerable overhead in terms of computation and memory allocation. In fact, to perform regression, each sensor node is required to keep a repository to store the received time information of the reference node. The capacity of this repository is determined in advance and it has a great impact on the estimation accuracy. Each time a new message is received, the most outdated received time information is deleted and the recently received information is stored at the repository. Then, least-squares regression is performed on the stored information to calculate the least-squares line. These calculation steps consist of many floating point multiplication and division operations, which are computationally expensive for sensor nodes \cite{Yildirim2012}. Moreover, recent studies  \cite{Lenzen2009Optimal,Yildirim2013} showed that slow-propagation of time information together with least-squares regression leads to exponentially increasing synchronization error with the network diameter, i.e. poor performance scalability. In order to improve synchronization performance and scalability, fast and reliable flooding is proposed in \cite{Lenzen2009Optimal}. However, the establishment of an efficient fast flooding architecture in WSNs is difficult \cite{FlashFlooding2009}. Recent solutions employ fragile hardware dependent software steps \cite{ferrari2011efficient}, demand different hardware capabilities \cite{kuo2012reconfiguring} and have serious scalability problems \cite{wang2012exploiting,wang2012triggercast}. As an alternative solution to the method of least-squares, which has several drawbacks in WSNs as we mentioned, the method of maximum likelihood estimation \cite{Chaudhari:10,Leng:11,Cao:13}, belief propagation \cite{Leng:11b}, and convex closure \cite{Berthaud:00} have also been proposed. However, these methods also have considerable computational and memory overheads and it is not clear whether they can be practically implemented in real WSNs since only simulative results are presented.

Differently from the flooding-based methods, in the \emph{fully-distributed} methods all sensor nodes interact only with their direct neighbours in a peer-to-peer fashion and there is no any special node which acts as a time reference for the other nodes. In mobile WSNs which are prone to frequent topological changes and node failures, \emph{fully-distributed} methods are the standard choice. Moreover, these methods are known to provide better synchronization among the neighbouring nodes since they share the so-called \textit{gradient property} \cite{Fan2006}, which is more appropriate for protocols such as TDMA. There are several fully distributed time synchronization protocols in the literature \cite{Firefly,Sommer2009Gradient,SchenatoFiorentin:2011,Choi:12}, however, they have fundamental shortcomings. The Reachback Firefly Algorithm (RFA)\cite{Firefly} can only provide synchronicity, hence it only ensures that sensor nodes agree on a firing period and phase but does not provide access to common notion of time. On the other hand, Average TimeSynch (ATS) \cite{SchenatoFiorentin:2011}, Gradient Time Synchronization Protocol (GTSP)\cite{Sommer2009Gradient} and Distributed Asynchronous Clock
Synchronization (DCS) \cite{Choi:12} require sensor nodes to store the time information of the neighbouring nodes for achieving distributed agreement. This requirement introduces a significant memory overhead in dense networks thus making infeasible for sensor nodes to keep track of all their neighbours due to their memory constraints. In fact, they suffer from the crucial problem of deciding which neighbours to keep track and which ones to discard, which decreases their scalability in dense networks. Moreover, their computation overhead is quite high and they require large amount of information to be exchanged among the neighbouring nodes. 

Finally, almost all protocols available in the literature, both in flooding-based methods and in fully distributed methods, and in particular those based on least-squares regression, cast the problem of time synchronization as a problem of the simultaneous estimation of relative clock drift and offset among two or more nodes, which give rise to time-synchronization algorithms  which are non-linear in the measurement noise. As a consequence, the effect of various error sources appear in the time synchronization error dynamics as multiplicative noise which makes the global time synchronization error to approximately grow exponentially with the diameter of the network, thus with poor performance scaling properties. Recently, a different approach based on control theory has been independently proposed in \cite{Chen:10} and \cite{Carli_2011} where synchronization is achieved by using linear feedback on the measured local synchronization error. The major advantage of this approach is that the  error sources appear as additive noise so that the global time synchronization error approximately grows as the square root of the network diameter.

\subsection{Contributions}

The existing time synchronization solutions for WSNs presented above suffer from two common drawbacks: high overhead in terms of computation and memory, and poor scalability in terms of global synchronization error. In this article, inspired by \cite{Carli_2011}, we propose and implement a control theoretic approach to time synchronization in WSNs which helps in eliminating these drawbacks.
In particular, we devise a distributed synchronization algorithm, named \textit{PISync}, which compensates the clock offsets and the differences in clock speeds based on a \textit{Proportional-Integral (PI) controller}. Roughly speaking,  nodes transmit their current time information periodically and upon receiving a synchronization message, they achieve time synchronization by applying a proportional feedback (P) and an integral feedback (I) on measured relative offsets  to compensate the differences between the clock offsets and the clock speeds respectively. PISync is linear, simple and easy to implement, which makes it suitable for both flooding-based and fully-distributed architectures. More precisely, we present specific time synchronization protocols based on PISync algorithm for the flooding-based scenario and for the fully distributed scenario. We provide simple design guidelines for the proportional (P)  and integral (I) feedback gains. We introduce an adaptive strategy for the integral gain in order to balance the trade-off between the convergence time and the steady state error. Moreover, we present a theoretical performance analysis of the proposed protocols and we show experimental comparisons with some popular time synchronization protocols, namely GTSP  \cite{Sommer2009Gradient}, FTSP \cite{Maroti2004} and PulseSync \cite{Lenzen2009Optimal}, based on a $5\times 4$ grid and on a 20-node line WSNs.

At the light of our theoretical analysis, experiments and simulations, the main advantages of the PISync protocols over the existing time synchronization protocols available in the WSNs literature, can be summarized as follows:
\begin{itemize}
\item \textbf{memory overhead}: all PISync protocols very small overhead in terms of main memory allocation, even in networks having very big densities. They do not store any distinct time information. 
\item \textbf{code footprint}: the code size of PISync protocols is very small, which provides remarkable gain for sensor nodes where flash memory (ROM) is a scarce resource.
\item \textbf{CPU overhead}: all PISync protocols are lightweight in terms of CPU usage. They have more than 97\% less computation overhead compared to the available protocols.
\item \textbf{blind communication}: all PISync protocols work in a completely blind manner and do not keep track of the neighbouring nodes. Hence, they are more robust to topological changes, node failures and new nodes joining into the network compared to existing approaches. This feature makes them also particularly suitable for mobile applications and for privacy-aware applications where nodes do not want to disclose their IDs.
\item \textbf{performance scalability:} all PISync protocols are scalable in terms of steady state global synchronization error, which grows with the square root of the network diameter.
\end{itemize}

\subsection{Organization of the Article}

The remainder of this article is organized as follows. In Section \ref{sec:System-Model}, we propose a system model for the clock of the nodes and the communication network in order to present and analyze our algorithms and protocols. We present a control theoretic analysis of the PISync algorithm and compare it with the classical least-squares based time synchronization in Section \ref{sec:PI}. In Section \ref{sec:Protocols}, we present and analyze fully distributed and flooding based time synchronization protocols based on PISync. Our implementation details, experimental evaluation and simulation results are presented in Section \ref{sec:Experiments}. Finally, conclusions and future research directions are presented in Section \ref{sec:Conclusion-and-Future}.

\section{System Model}

\label{sec:System-Model}

In this section, we propose a hardware clock model, a logical clock model and a network model which we use for the presentation and analysis of the synchronization algorithms in this article.

\subsection{Hardware Clock Model}

We start by proposing a model of each built-in \textbf{hardware clock}, i.e. local clock, of the sensor nodes which is simple enough but which captures the main difficulty of the synchronization problem, namely, the fact that the time is an unknown variable, which has to be estimated. 

\begin{figure}
\begin{center}
\begin{tabular}{c}
\includegraphics[width=0.5\columnwidth]{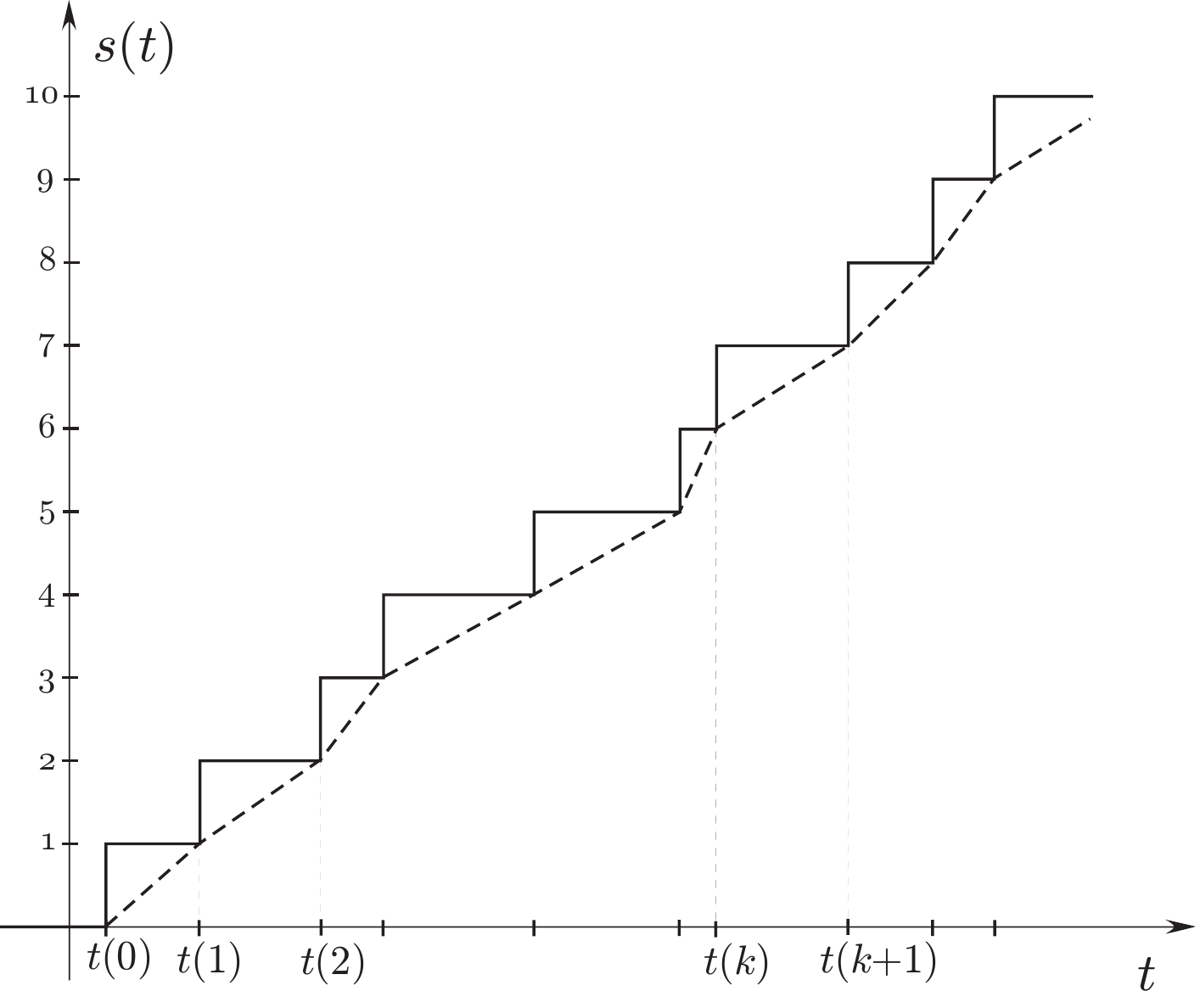}
\end{tabular}
\caption{\label{fig:sdit} The graphs of $s(t)$ (continuous line) and of its approximation $\int_{-\infty}^t f(\sigma)d\sigma$ (dashed line).  Due to their dynamic frequencies, the tick events of clocks do not occur at regular time intervals. Hence, $s(t)$ is a step shaped function.}
\end{center}
\end{figure}

Assume that each hardware clock has an oscillator capable to produce an event at time $t(k)$, $k\in \mathbb{N}$. The hardware clock has to use these ticks in order to estimate the time. The following cumulative function well describes the time evolution of the tick counter that can be implemented in the hardware clock
$$
s(t)=\sum_{k=0}^{\infty} {\bf{1}}(t-t(k))
$$  
where ${\bf{1}}$ is the unit step function, i.e.,
$$
{\bf{1}}(t)=
\left\{
\begin{array}{lc}
0 & t\le 0\\
1& t> 0
\end{array}
\right.
$$
In this way the counter output is the step shaped function shown in Figure \ref{fig:sdit}.
Notice that the function
\begin{equation}\label{eq:regular_f}
f(t):=\frac{1}{t(k+1)-t(k)}\qquad\mbox{for all $t\in[t(k),t(k+1)[$},
\end{equation}
can be interpreted as the oscillator frequency at time $t$ and that $s(t)\simeq\int_{-\infty}^t f(\sigma)d\sigma$. Typically a nominal value $\hat f$ of $f(t)$ is known together with a lower bound  
$f_{min}$ and an upper bound $f_{max}$ such that $f(t)\in[f_{min},f_{max}]=[\hat{f}-\Delta\! f_{max},\hat{f}+\Delta\! f_{max}]$ where $\Delta\! f_{max}=\frac{f_{max}-f_{min}}{2}$.

\subsection{Logical Clock Model}

Collecting time information from the neighboring nodes, each node calculates a logical clock which represents the network-wide global time. Let $t_0$ be the latest time at which the logical clock has been updated. From the counter, one can build a time estimate $\hat{t}(t)$ at any time $t>t_0$ by letting
\begin{equation}
  \label{eq:hat_t}
  \hat{t}(t) = \hat{t}(t_0) + \hat{\Delta}(t)[s(t)-s(t_0)], 
\end{equation}
where $\hat{\Delta}(t)$ is an estimate of the oscillation period $1/f(t)$ in the period $[t_0, t]$. It is reasonable to initialize $\hat{\Delta}(t)$ to $1/\hat{f}$. The time estimate $\hat{t}(t)$ can be considered as the value of the \textbf{logical clock} at time $t$ and represents the network-wide global time. It can be observed that the estimate  $\hat{\Delta}(t)$, which we will also refer to as the \textbf{rate multiplier}, represents the progress rate (speed) of the logical clock.

Both the logical clock value $\hat{t}(t)$ and its progress rate $\hat{\Delta}(t)$ can be modified when the node obtains information allowing it to improve its time and oscillator frequency estimates. Assume that these corrections are applied at time instants $T_\mathrm{up}(h)$, where $h=0,1,\dots$, called updating time instants. In this case we have \footnote{Given the time $t$, with the symbol $t^+$ we mean the time instant just after $t$.}
\begin{align*}
    \hat{t}(T_\mathrm{up}^+(h))& = \hat{t}(T_\mathrm{up}(h)) + u'(h) \\
    \hat{\Delta}(T_\mathrm{up}^+(h))& = \hat{\Delta}(T_\mathrm{up}(h)) + u''(h)
\end{align*}
where $u'$ and $u''$ denote the control inputs applied to $\hat{t}$ and $\hat{\Delta}$, respectively. For $t\in\left(T_\mathrm{up}^+(h), T_\mathrm{up}(h+1)\right)$, $ \hat{t}(t)$ is updated according to \eqref{eq:hat_t}, while $\hat{\Delta}$ is left unchanged. Specifically,
\begin{align}
  \label{eq:Update}
    \hat{t}(t)& = \hat{t}(T_\mathrm{up}^+(h)) + \hat{\Delta}(T_\mathrm{up}^+(h)) \left(s(t)-s(T_\mathrm{up}(h))\right), \ \ \ t\in\left(T_\mathrm{up}^+(h), T_\mathrm{up}(h+1)\right)
\nonumber \\
    \hat{\Delta}(t)&=\hat{\Delta}(T_\mathrm{up}^+(h))\end{align}

\subsection{Network Model}

We represent a WSN  by a graph $G=(V,\mathcal{E})$ where the vertex set $V=\left\{1,\ldots,N\right\}$ represents sensor nodes. For each node $i\in\left\{1,\ldots,N\right\}$, let $f_i(t)$ be the evolution of the oscillator frequency of the hardware clock of that node. Moreover, for $i\in\left\{1,\ldots,N\right\}$, let $[\hat{t}_i(t)\ \hat{\Delta}_i(t)]$ denote the state of the logical clock of node $i$. We assume that the nodes can exchange the states of their logical clocks according to a graph $G$ where $(i,j)\in \mathcal{E}$ whenever the node $i$ can send its logical clock state to the node $j$.  
 
Specifically, each node $i\in \left\{1,\ldots,N\right\}$, broadcasts its estimate $\hat{t}_i(t)$, i.e., the value of its logical clock, to its neighbours at time instants $T_{\mathrm{tx},i}(h)$, $h=0,1,\dots$, and can use any information it receives from the neighbouring nodes to apply a control at the time instants $T_{\mathrm{up},i}(h)$, $h=0,1,\dots$.

\begin{figure}[!t]
\centering
\includegraphics[scale=0.35]{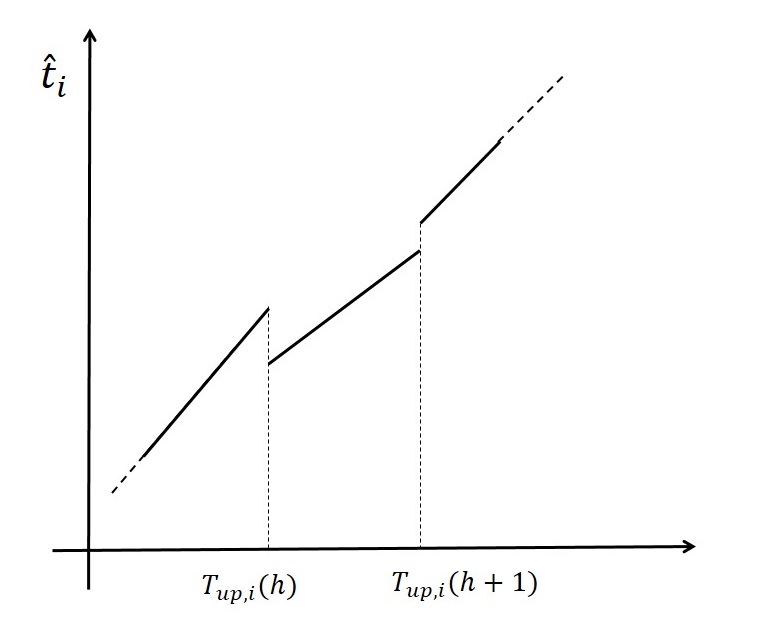}\,\,\,\,\includegraphics[scale=0.35]{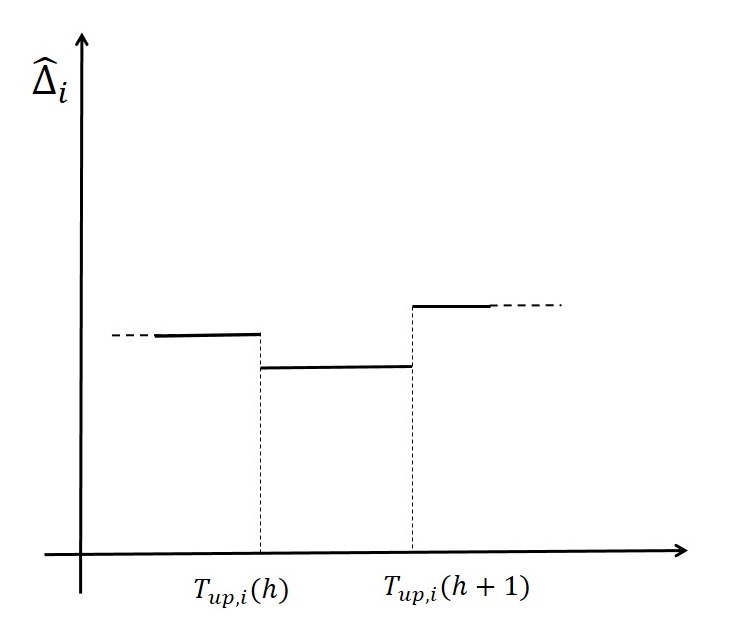}
\caption{Behavior of the logical clock $\hat{t}_i$ ({\bf left panel}). Behavior of the rate multiplier $\hat{\Delta}_i$ ({\bf right panel}).
}
\label{fig:Updating}
\end{figure}

More precisely
\begin{align}
  \label{eq:Update_i}
  \hat{t}_i(T_{\mathrm{up},i}^+(h)) &= \hat{t}_i(T_{\mathrm{up},i}(h))+ u'_i(h)\nonumber\\
\hat{\Delta}_i(T_{\mathrm{up},i}^+(h))& = \hat{\Delta}_i(T_{\mathrm{up},i}(h))+ u''_i(h)
\end{align}
where $[u'_i(h)\ u''_i(h)]$ is the control action applied at time $T_{\mathrm{up},i}(h)$.
Moreover for $t\in\left[T_{\mathrm{up},i}^+(h), T_{\mathrm{up},i}(h+1)\right]$ we assume that the state $[\hat{t}_i(t)\ \hat{\Delta}_i(t)]$ is updated according to~\eqref{eq:Update}. In Figure \ref{fig:Updating} we depict the behaviour of the logical clock $\hat{t}_i$ and of the rate multiplier $\hat{\Delta}_i$.

\section{PI Synchronization (PISync) Algorithm }
\label{sec:PI}

In this section, we devise a synchronization algorithm suitable for WSNs, named PISync, which synchronizes the logical clocks of the nodes based on a Proportional-Integral (PI) Controller. 
To make easier the description of the algorithm and of its stability properties we assume in the Section that the oscillator frequencies $f_i$ are constant, i.e., for $i\in\left\{1,\ldots,N\right\}$, $f_i(t) = \bar{f}_i$ for all $t\in\mathbb{R}_{>0}$. 

\vspace{0.1cm}

From \eqref{eq:hat_t}, it follows that~\eqref{eq:Update}, for the $i$-th logical clock, can be equivalently rewritten as
\begin{align}\label{eq:UpdateDelayed-f_constant}
    \hat{t}_i(t) & =  \hat{t}_i(T_{\mathrm{up},i}^+(h)) + \hat{\Delta}_i(T_{\mathrm{up},i}^+(h))\,\bar{f}_i\,(t-T_{\mathrm{up},i}(h))\nonumber\\
  \hat{\Delta}_i(t) & =  \hat{\Delta}_i(T_{\mathrm{up},i}^+(h)).
\end{align}
 
 The objective is to find a control strategy yielding the clock synchronization, namely such that there exist constants $a\in\mathbb{R}_{>0}$ and $b\in\mathbb{R}$ such that synchronization errors 
\begin{equation}
  \label{eq:SynchErrors}
  e_i(t) := \hat{t}_i(t) - (a t + b), \quad i \in \left\{1,\ldots,N\right\},
\end{equation}
converge  to zero or remain small. In other words, the objective is to synchronize the variables $\hat{t}_i(t)$, $i\in\left\{1,\ldots,N\right\}$, namely, to find a law which allows the logical clocks of the nodes to obtain the same time estimate.

\begin{figure}
\begin{center}
\begin{tabular}{c}
\includegraphics[width=0.6\columnwidth]{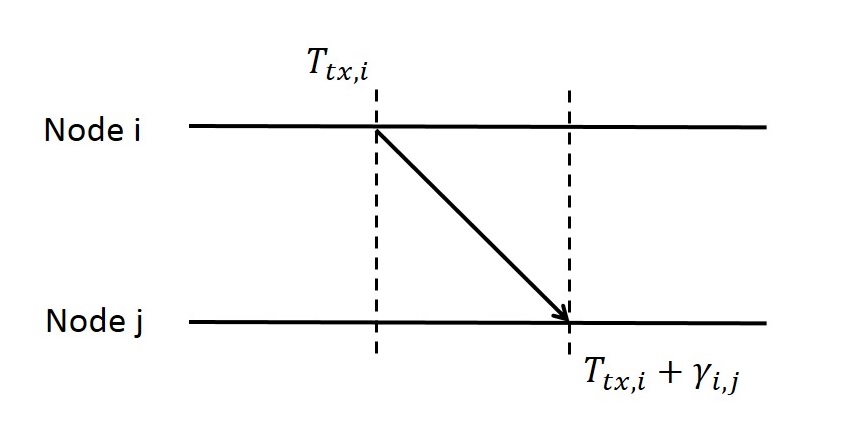}
\end{tabular}
\caption{\label{fig:tx}The transmission delay during the communication between nodes i and j.}
\end{center}
\end{figure}

Now, without loss of generality, assume that node $i$ transmits, at a generic transmission time $T_{\mathrm{tx},i}$, the information $\hat{t}_i(T_{\mathrm{tx},i})$ to node $j$. Due to transmission delays, 
the information $\hat{t}_i(T_{\mathrm{tx},i})$ is received by node $j$ at a delayed time
$$
T_{\mathrm{tx},i}+\gamma_{i,j}
$$
where $\gamma_{i,j}$ is a non-negative real number representing the deliver delay between $i$ and $j$ (see Figure \ref{fig:tx}).

Based on the information received, node $j$ instantaneously applies to its current state $[\hat{t}_j(T_{\mathrm{tx},i}+\gamma_{i,j})\,\,\,\,  \hat{\Delta}_j(T_{\mathrm{tx},i}+\gamma_{i,j})]$ the following correction
\begin{align}\label{eq:control-u}
u'_j &=   \beta_j \cdot \left(\, \hat{t}_i(T_{\mathrm{tx},i})-\hat{t}_j \left(T_{\mathrm{tx},i}+\gamma_{i,j}\right)\,\right)\nonumber\\
u''_j &=   \alpha_j \cdot \left(\, \hat{t}_i(T_{\mathrm{tx},i})-\hat{t}_j \left(T_{\mathrm{tx},i}+\gamma_{i,j}\right)\,\right)
\end{align}
where $\beta_j$, $\alpha_j$ are two control parameters to be designed.

Observe that, according to the above model, we have that $T_{\mathrm{up},j}=T_{\mathrm{tx},i}+\gamma_{i,j}$, therefore we can write:
\begin{align}
  \label{eq:Update_i1}
  \hat{t}_j\left(\,\left(T_{\mathrm{tx},i}+\gamma_{i,j}\right)^+\,\right) &= \hat{t}_j\left(\,T_{\mathrm{tx},i}+\gamma_{i,j}\,\right)+ \beta_j \cdot \left(\, \hat{t}_i(T_{\mathrm{tx},i})-\hat{t}_j \left(T_{\mathrm{tx},i}+\gamma_{i,j}\right)\,\right)\nonumber\\
\hat{\Delta}_j\left(\,\left(T_{\mathrm{tx},i}+\gamma_{i,j}\right)^+\,\right) & = \hat{\Delta}_j\left(\,T_{\mathrm{tx},i}+\gamma_{i,j}\,\right) +  \alpha_j \cdot \left(\, \hat{t}_i(T_{\mathrm{tx},i})-\hat{t}_j \left(T_{\mathrm{tx},i}+\gamma_{i,j}\right)\,\right)
\end{align}

The update rule above indicates that only the \textit{estimated synchronization error} $ \hat{t}_i(T_{\mathrm{tx},i}(h))-\hat{t}_j \left(T_{\mathrm{up},j}(h')\right)$ between nodes $i$ and $j$, and the two control parameters $\alpha_i,\beta_i$ are used to update the value and the rate multiplier of the logical clock of node $i$. Hence, PISync does not require any extra information.

\begin{remark}
Observe that the control law above introduced can be seen as a PI controller where $u'_j$ and $u''_j$ represent, respectively, the proportional and the integral part. It is worth mentioning that this strategy has been inspired by the PI consensus controller strategy proposed in \cite{Carli_2011}. However in \cite{Carli_2011} the synchronization protocol is analyzed under the following assumptions:
\begin{enumerate}
\item all the nodes perform the transmitting and the updating actions synchronously (\emph{synchronous implementation}), i.e., $T_{\mathrm{tx},i}(h)=T_{\mathrm{up},i}(h)=hT$ for all $i \in \left\{1,\ldots,N\right\}$ and for all $h=0,1,\ldots$, where $T$ is a pre-assigned sampling time;
\item the control parameters $\alpha$ and $\beta$ are assumed constant;
\item the hardware frequencies $\bar{f}_i$, $i \in \left\{1,\ldots,N\right\}$ are quite close to each other.
\end{enumerate}

Under the above assumptions, the authors in \cite{Carli_2011} provided sufficient conditions on $\beta$ and $\alpha$ which guarantee that the synchronization is reached. Specifically they showed that it must hold $0< \beta \leq 1$ and $0 < \alpha < \bar{\alpha}$ where $\bar{\alpha}$ depends on some properties of the Laplacian matrix associated to the graph $\mathcal{G}$, which can be calculated only having a global knowledge of the network. A similar control oriented approach has been proposed in \cite{Chen:10} where the parameters $K_I$ and $K_P$ provided in that work  play the role of $\alpha$ and $\beta$ here. However, the design guidelines for  $K_I$ and $K_P$ are derived by considering only two nodes, therefore as shown in \cite{Carli_2011}, they might give rise to unstable behaviours of the algorithm for large scale networks. 

The goal of this paper is to extend the control-based synchronization strategy with PI feedback to the more realistic scenarios where the nodes do not communicate synchronously with each other, and to be suitable to time-varying scenarios where new nodes can be added or existing nodes can die and move. In particular we endow each node with adaptive laws for  the real time design of the parameters $\alpha$ and $\beta$ based only on the information coming from their neighbouring nodes. Moreover, the design strategy for $\alpha$ and $\beta$ provided in this work, is independent of the network topology and revolves the potential instability problem present in the protocol suggested in \cite{Chen:10}.

\end{remark}

\subsection{Pairwise control-based time synchronization: rate of converge}
\label{subsec:pairwise_control}

Before providing some design guidelines for the parameters $\alpha$ and $\beta$ we would like to give an intuitive explanation of the proposed control-based solution for time synchronization and highlight the major differences with the standard solutions available in the literature based on least-squares. To this purpose, we assume there are only two nodes $i$ and $j$, where $i$ plays the role of the reference clock. Moreover, without loss of generality, we assume that node $i$ is perfectly synchronized with respect to the absolute time, i.e. $\bar f_i=\hat f$ and $\hat t_i(0)=0$, therefore Eqn.\eqref{eq:SynchErrors} becomes
$$ e_j(t)=\hat{t}_j(t)-\hat{t}_i(t)=\hat{t}_j(t)-\left(\frac{\overline{f}_i}{\hat f}t+\hat{t}_i(0)\right)=\hat{t}_j(t)-t.$$
We also assume that the reference node $i$ periodically with period $B$ transmits a message with its own logical clock, which in this case coincides with the hardware clock, and that there is no delay, i.e. $\gamma_{ji}=0$, therefore we have
$$ T_{tx,i}(h)=hB, \ \ h\in \mathbb{N}$$

In the context of \textbf{control-based time synchronization}, with a little abuse of notation, we denote $\hat t_j(h)=\hat t_j(hB)$ and similarly for $\hat t_i$ and $e_j$, therefore the update in Eqn.\eqref{eq:Update_i1} becomes
\begin{align}
  \label{eq:up_simp}
\hat{t}_j(h^+) &=\hat{t}_j(h) - \beta_j e_j(h)\\
\hat{\Delta}_j(h^+)& = \hat{\Delta}_j(h)- \alpha_j e_j(h).\end{align}
By substituting the previous equations into Eqn.~\ref{eq:UpdateDelayed-f_constant} with $t=(h+1)B$ and by recalling the definition of $e_j$ and that $\hat t_i(h+1)=\hat t_i(h)+ B$, we obtain a state space description of the error dynamics:

\begin{equation}\label{eqn:state_space}
\left[ \begin{array}{c} e_j(h+1) \\ \hat \Delta_j(h+1) \end{array} \right] =\underbrace{\left[ \begin{array}{cc} 1-\beta_j -\alpha_j B \bar f_j& B \bar f_j\\ -\alpha_j & 1\end{array} \right]}_{F} \left[ \begin{array}{c} e_j(h) \\ \hat \Delta_j(h) \end{array} \right]-\left[ \begin{array}{c}  B \\ 0 \end{array} \right] , \ \ \ \left[ \begin{array}{c} e_j(0) \\ \hat \Delta_j(0) \end{array} \right] =\left[ \begin{array}{c} \hat t_j(0) \\  \frac{1}{\hat f} \end{array} \right]   
\end{equation}
where $\bar{f}_j$ and $\hat t_j(0)$ are arbitrary. By inspecting these dynamics, it is immediate to see that the dynamics of $\hat \Delta_j$ is a discrete time integrator driven by the error signal $e_j$ where the parameter $\alpha$ is referred as the \emph{integrator feedback gain}. The dynamics of the synchronization error $e_j$ are affected by the output of the integrator $\hat \Delta_j$ and includes also a proportional feedback on itself via the \emph{proportional feedback gain} $\beta$. If the gains $\alpha$ and $\beta$ are chosen such that the matrix $F$ is strictly stable, i.e. its two eigenvalues have modulus strictly smaller than unity, then the dynamical system must converge asymptotically to a steady state, i.e. $\lim_{h\to \infty}e_j(h)=e_j(\infty)$ and $\lim_{h\to \infty}\hat \Delta_j(h)=\hat \Delta_j(\infty)$ which must satisfy the following expressions:
$$F \mbox{ strict. stable} \Longrightarrow \left\{\begin{array}{l} \hat{\Delta}_j(\infty)=  \hat{\Delta}_j(\infty) -\alpha_j e_j(\infty) \Longrightarrow e_j(\infty)=0, \ \ (\alpha_j\neq 0)\\
e_j(\infty) = (1-\beta_j -\alpha_j B \bar f_j) e_j(\infty) +B \bar f_j \hat{\Delta}_j(\infty) - B \Longrightarrow \hat{\Delta}_j(\infty) =\frac{1}{\bar f_j}\end{array}\right.$$
which shows that time synchronization is eventually achieved, since $e_j(\infty)=0$. The role of the proportional gain is to compensate for the different initial clock offset $\hat t_j(0)$, while the role of the integrator gain is to compensate for different clock drifts $\bar f_j-\hat f$, which are both not directly measurable. In fact, if the integrator is disabled and the proportional gain is chosen to keep the dynamics stable, then the a steady state error is present:
$$ \alpha_j=0 \Longrightarrow \left\{\begin{array}{l}  \hat{\Delta}_j(h)=\frac{1}{\hat f}, \ \ \  e_j(h+1) = (1-\beta_j)e_j(h)+B\frac{\bar f_j-\hat f}{\hat f} \\ e_j(\infty) =   \frac{B(\bar f_j-\hat f)}{\beta_j\hat f}, \ \ \ (0<\beta_j<2)\end{array}\right.$$
Note that the steady state error is directly proportional to both the difference of the relative clock speed and the transmission period $B$. In fact, if all clocks have the same drift, i.e. $\bar f_j=\hat f, \forall j$, then the proportional feedback alone would suffice to drive the synchronization error to zero. 

As mentioned above, the PI synchronization strategy achieves synchronization only if the matrix $F$ is stable. Moreover, linear dynamical system theory also shows that the rate of convergence is given by the eigenvalues of such matrix. The eigenvalues of the matrix $F$ are given by the solution of the following second order systems:
$$ z^2-(2-\beta_jl-\alpha_j B \bar f_j)z+(1-\beta_j)=0\Longrightarrow z_{1,2}= \frac{2-\beta_j-\alpha_j B \bar f_j\pm \sqrt{(\beta_j+\alpha_j B \bar f_j)^2-4\alpha_j\bar f_j B}}{2}.$$
We then would like to find the set of values for the gains $\alpha_j$ and $\beta_j$ for which these roots are stable, i.e., $|z_{1,2}|<1$. After some simple calculations, we obtain:
\begin{equation} \label{eqn:alpha_beta}
0<\beta_j < 2, \ \  0<\alpha_j < \frac{2(2-\beta_j)}{\bar f_j B} 
 \end{equation}
which shows that if $\alpha_j$ is positive but sufficiently small, then synchronization is achieved.

 It is also possible to find the optimal value for $\alpha_j$ for any fixed and feasible value of $\beta_j$ to maximize the rate of convergence, which after some simple calculations is given by:
\begin{equation}\label{eqn:alpha_star}
\alpha_j^* = \mathrm{argmin}_{\alpha_j} |z_{1,2}|=\frac{2-\beta_j}{\bar f_j B}, \ \ \ |z_{1,2}(\alpha_j^*)|=\sqrt{|1-\beta_j|}, \ \ 0<\beta_j<2.
\end{equation}
From this expression it is clear that the integrator gain $\alpha_j$ should be decreased if the communication period $B$ is increased, in order to maintain the same eigenvalues of the matrix $F$, i.e., if the same rate of convergence is desired. 
It is also of interest, as shown in the next section, to compute the rate of convergence for the special case $\beta_j=1$, which gives
\begin{equation}\label{eqn:rho_alpha} \beta_j=1 \Longrightarrow z_1 = 0, z_2 = 1-\alpha \hat f B
\end{equation}

\begin{figure} 
\center

\includegraphics[scale=0.4]{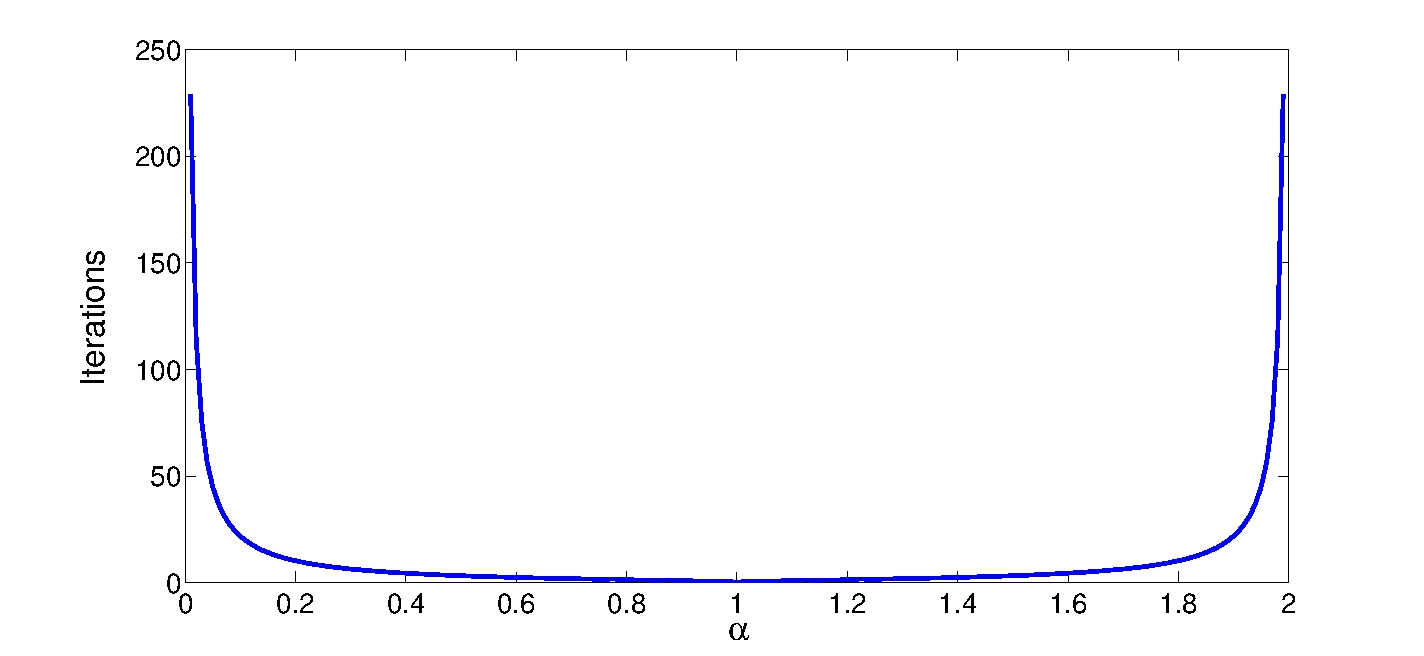} 

\caption{\label{fig:alphaVSiterations} Plot of the trade-off between the value of $\alpha$ (rescaled by $\hat f B$) and the number of iterations required by the algorithm to reduce the error of 90\% of its initial value. The plot has been obtained assuming $\beta_j=1$, $B=30$sec and $\hat f =1MHz$.}
\end{figure}

In Figure \ref{fig:alphaVSiterations} we plotted the trade-off between the value of $\alpha$ (rescaled by $\hat f B$) and the number of iterations required by the algorithm to reduce the error of 90\% of its initial value.

\subsection{Pairwise control-based time synchronization: steady-state error}
\label{sec:noise2}

When dealing with wireless devices, there are several noise sources, like quantization effects, presence of unreliable communication channels, measurements errors, time-varying oscillator frequencies, transmission delays, which prevent any synchronization algorithm from reaching the exact network synchronization. 

In this section we consider the presence of transmission noise and time-varying frequencies and we compute their effect in the steady state error. Specifically, assume that node $i$ transmits, at time $T_{\mathrm{tx},i}(h)$, the information $\hat{t}_i(T_{\mathrm{tx},i}(h))$ to node $j$. Due to quantization effects and in the presence of unreliable channels, 
the information received by node $j$ can be conveniently modelled as
$$
\hat{t}_i(T_{\mathrm{tx},i}(h))+v_{i,j}(T_{\mathrm{tx},i}(h))
$$
where $v_{i,j}(T_{\mathrm{tx},i}(h))$ is a zero mean white noise of variance $r$, i.e.,$\mathbb{E}\left[v_{i,j}(T_{\mathrm{tx},i}(h))\right]=0$ and $$\mathbb{E}\left[v^2_{i,j}(T_{\mathrm{tx},i}(h))\right]=\sigma_v^2=\eta^2_t \frac{1}{\hat f^2}$$
where $\eta_t$ is an adimensional parameters which is typically in the order of unity.
The standing assumption is that noises related to different transmissions are independent from each other.  Additionally we assume the frequency of a generic node $k$ at time $T_{\mathrm{up},k}(h)$ to be given by 
$$
\bar{f}_k+w_k(h)
$$
where $w_k(h)$ is a zero mean-noise uniformly distributed in $\left[-\Delta f_{max}, \Delta f_{max}\right]$ with corresponding variance $$\sigma_w^2=\frac{(\Delta f_{max})^2}{12}= \eta_w^2 \hat f^2$$
where $ \eta_f$ is an adimensional parameter which refers to the typical relative frequency change over the course of one synchronization period $B$.

Now, without loss of generality, let us assume that node $1$ is the reference node and let node $2$ be a child of node $1$. Next, we provide a mean-square analysis to characterize the synchronization error between node $2$ and node $1$. 
In our analysis we assume that $T_{\mathrm{tx},1}(h)=hB$, where $B$ is a given sampling time and $h=0,1,2,\ldots$. Moreover we assume that the transmission delays are negligible, thus $T_{\mathrm{up},2}(h)=T_{\mathrm{tx},1}(h)$. For the sake of the notational convenience let $\hat{t}_i(h)=\hat{t}_i(T_{\mathrm{up},i}(h))$, $\hat{t}_i(h^+)=\hat{t}_i(T_{\mathrm{up},i}^+(h))$, $\hat{\Delta}_i(h)= \hat{\Delta}_i(T_{\mathrm{up},i}(h))$, $\hat{\Delta}_i(h^+)= \hat{\Delta}_i(T_{\mathrm{up},i}^+(h))$, for $i \in \left\{1,2\right\}$. For simplicity assume that $\hat{\Delta}_1(h)=1/\hat{f}$ for all $h$. We have that
\begin{eqnarray*}
    \hat{t}_1(h+1) & = & \hat{t}_1(h) +\frac{1}{\hat{f}}(\bar{f}_1+w_1(h))\,B
\end{eqnarray*}
and 
\begin{eqnarray*}
    \hat{t}_2(h+1) & = & \hat{t}_2(h^+) + \hat{\Delta}_2(h^+)\,(\bar{f}_2+w_2(h))\,B\nonumber\\
  \hat{\Delta}_2(h+1) & = & \hat{\Delta}_2(h^+)
\end{eqnarray*}
where
\begin{align}
\hat{t}_2(h^+) &=\hat{t}_2(h) - \beta_2 (e_2(h)-v_{1,2}(h))\\
\hat{\Delta}_2(h^+)& = \hat{\Delta}_2(h)- \alpha_2 (e_2(h)-v_{1,2}(h)).
\end{align}
We further consider the approximation $\bar{f}_1=\bar{f}_2=\hat f$ since the true frequencies are very close to the nominal value $\hat f$. Finally, we set $\beta_2=1$. The reason for choosing $\beta_2=1$ comes from the observation provided in the previous section that with $\alpha=\frac{1}{B\hat f}$, then the time of convergence is minimized, and we interested in computing the steady state error as a function of $\alpha$.
Recalling that $e_2(h)=\hat{t}_2(h)-\hat{t}_1(h)$, and by defining $z_2(h)=\hat{\Delta}_2(h)\hat f_2-1$, then after some straightforward calculation we get: 

\begin{align*}
e_2(h+1)
&=v_{1,2}(h)+ B\left(1+w_2(h)\frac{1}{\hat f }\right)z_2(h+1)+\frac{B}{\hat f }(w_2(h)-w_1(h))
\end{align*}
and
\begin{align*}
z_2(h+1)&= z_2(h)-\alpha \hat f(e_2(h)-v_{1,2}(h))\\
&=z_2(h)-\alpha  \hat f \left(v_{1,2}(h-1)+ B\left(1+w_2(h-1)\frac{1}{\hat f}\right)z_2(h)+\frac{B}{\hat f}(w_2(h-1) -w_1(h-1))-v_{1,2}(h)\right)\\
&= \left(1-\alpha  \hat f B -\alpha w_2(h-1)B\right)z_2(h)-\alpha \hat f \left(v_{1,2}(h-1)+\frac{B}{\hat f}(w_2(h-1) -w_1(h-1))-v_{1,2}(h)\right)
\end{align*}
Let $P_{z_2}(h)=\mathbb{E} \left[z^2_2(h)\right]$ and $P_{e_2}(h)=\mathbb{E} \left[e^2_2(h)\right]$. It follows
\begin{align*}
P_{z_2}(h+1)&= \left((1-\alpha \hat f  B)^2 +\alpha^2 \sigma^2_w B^2 \right)P_{z_2}(h) +\alpha^2\left(2\sigma^2_v  \hat f^2+ 2\sigma_w^2B^2\right)
\end{align*}
Hence
$$
\lim_{h \to \infty} P_{z_2}(h) = \frac{2 \alpha \left(\eta_t^2 +\eta_w^2 \hat f^2 B^2 \right)}{2 B \hat f -\alpha B^2 \hat f^2(1+\eta_w^2)}
$$
and, in turn,
\begin{equation}\label{eq:AsympError}
\bar{P}_{e_2} =\lim_{h \to \infty} P_{e_2}(h) = \frac{2 \alpha B^2\left(\eta_t^2 +\eta_w^2 \hat f^2 B^2 \right)\left(1+\eta^2_w\right)}{2 B\hat f -\alpha B^2 \hat f^2(1+\eta_w^2)}+\eta_t^2\frac{1}{\hat f^2}+ 2\eta_w^2 B^2
\end{equation}
From the previous expression it is clear the error is an increasing function of $\alpha$ and that even for $\alpha=0$ there is still some non-zero steady state error.

\begin{figure} 
\center
\includegraphics[scale=0.3]{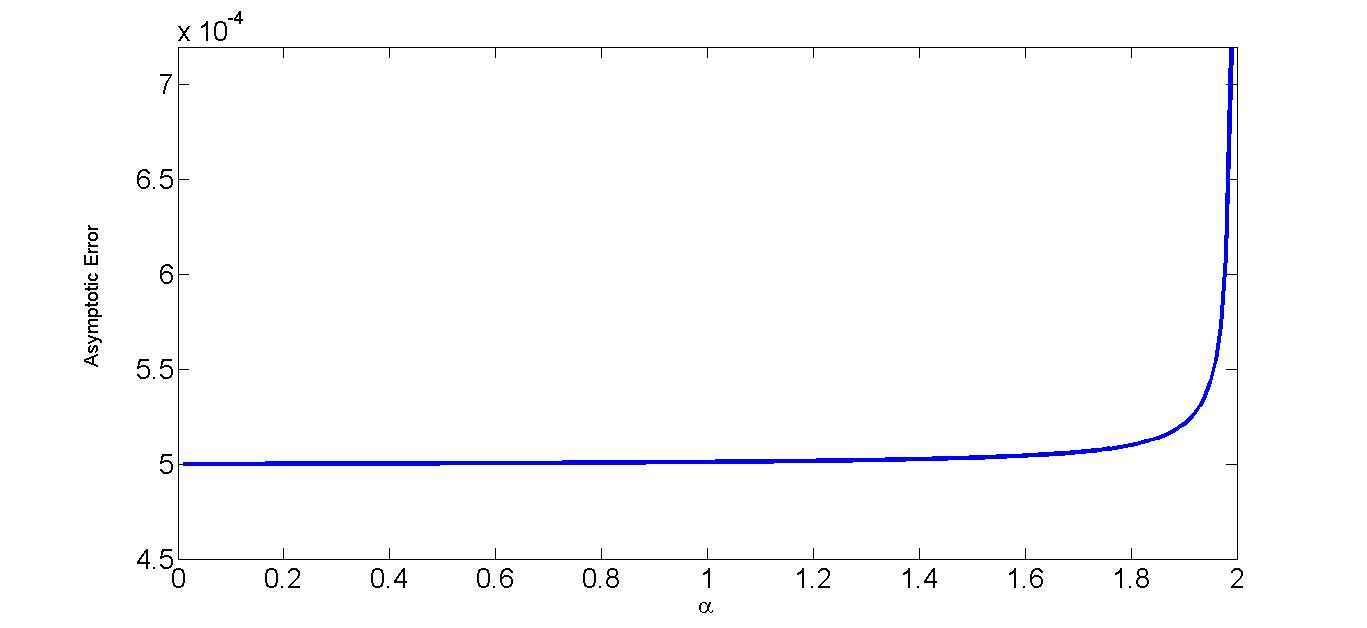} 
\caption{\label{fig:errorVSalpha} Plot of the trade-off between the mean-square asymptotic error and the value of $\alpha$ (re-scaled by $\hat f B$). The plot has been obtained assuming $\beta_j=1$, $B=30$sec, $\hat f =1MHz$, $\sigma_v^2=5*10^{-4}$sec and $\sigma_w^2=10^{-8}$.}
\end{figure}

In Figure \ref{fig:errorVSalpha}  we plotted the trade-off between the mean-square asymptotic error and the value of $\alpha$ (re-scaled by $\hat f B$). One can see that the smaller the value of $\alpha$ is, the smaller the value of the error is. However, as seen in the previous Figure, as $\alpha$ approaches $0$, the number of iterations required to converge becomes larger and larger.

\subsection{Control-based vs least-squares-based time synchronization}
\label{subsec:control_vs_least_squares}

Under the same hypothesis, we now would like to provide a similar intuitive presentation of the \textbf{least-squares-based time synchronization}.  In this context, the equations for the logical clocks of the two nodes can be written as:
$$ \hat{t}_i(h) =\frac{1}{\hat f}s_i(h) $$
$$ \hat{t}_j(h) = u_j'(h)+u_j''(h)s_j(h) $$
where $u_j'(h)$ and $u_j''(h)$  have to be designed to drive the synchronization error $e_j(h)=e_j(h)-e_i(h)$ to zero. We assume that $u'_j(h)=u'$ and $u''_j(h)=u''$ are kept constant for the first $H$ steps so that we can write
$$ \underbrace{\left[\begin{array}{c} e_j(H-1) \\ \vdots \\ e_j(0)  \end{array}\right]}_{\mathbf{e}}=\underbrace{\left[\begin{array}{cc} 1 & s_j(H-1) \\ \vdots & \vdots \\ 1 & s_j(0)  \end{array}\right]}_{A} \underbrace{\left[\begin{array}{c} u' \\ u''  \end{array}\right]}_{\mathbf{u}}- \underbrace{\frac{1}{\hat f }\left[\begin{array}{c} s_i(H-1) \\ \vdots \\ s_i(0)  \end{array}\right]}_{\mathbf{b}}$$
Under least-square-based synchronization the values for the compensating parameters is given by
$$ \mathrm{argmin}_{\mathbf{u}} \|\mathbf{e}\| = \mathrm{argmin}_{\mathbf{u}}  \| A\mathbf{u}-\mathbf{b}\|\Longrightarrow \mathbf{u}=(A^TA)^{-1}A^T\mathbf{b}$$
Note that $A$ and $b$ are known to node $j$ as long as node $i$ transmits either $\hat t_i(h)$ or $s_i(t)$ since $\hat f $ is known. In the specific case when $H=2$, the solution is given by:
\begin{equation} \label{eqn:LS}
u'= \frac{1}{\hat f }\frac{s_i(0)s_j(1)-s_i(1)s_j(0)}{s_j(1)-s_j(0)}, \ \ \ \ u'' = \frac{1}{\hat f }\frac{s_i(1)-s_i(0)}{s_j(1)-s_j(0)}.
\end{equation}
If no measurements errors are considered, then the previous solutions provide exact synchronization, i.e., $\hat t_j(h)=\hat t_i (h), \forall h\geq 2$. In practice, this is not the case due, for example, to transmission delay or quantization, and therefore the previous procedure has to be repeated periodically. 

At the light of the derivations above about the control-based and least-squares-based time synchronization a number of observations are in order:
\begin{itemize}
\item Under the ideal scenarios above, the rate of convergence for the control-based synchronization is asymptotic, while using least-squares can be achieved in finite time. The better performance in terms of convergence rate for the least-squares-based methods continues to hold also in the more general WSNs scenarios as shown in Section \ref{sec:Experiments}.
\item If measurements errors are present, due for example to transmission delay or quantization, then we have to substitute  $\hat{t}_{i}(h)\leftarrow \hat{t}_{i}(h)+w_{i,j}(h)$ into Eqn.~\eqref{eqn:state_space} and $s_{i}(h)\leftarrow s_{i}(h)+w_{i,j}(h)$ into Eqn.~\eqref{eqn:LS}, where $w_{i,j}(h)$'s represent the measurement noise at iteration $h$. As the consequence the time synchronization error dynamics for the control-based strategy becomes:
\begin{equation}\label{eqn:noise}
 \left[ \begin{array}{c} e_j(h+1) \\ \hat \Delta_j(h+1) \end{array} \right] =\left[ \begin{array}{cc} 1-\beta_j & B \bar f_j\\ -\alpha_j & 1\end{array} \right]\left[ \begin{array}{c} e_j(h) \\ \hat \Delta_j(h) \end{array} \right]-\left[ \begin{array}{c}  B \\ 0 \end{array} \right] +w_{i,j}(h)  \left[ \begin{array}{c} \beta_j \\ \alpha_j \end{array} \right]  \end{equation}
while the equation $u''$ for the first iteration in the least-squares approach becomes
$$ u'' = \frac{1}{\hat f }\frac{s_i(1)-s_i(0)+w_{i,j}(1)-w_{i,j}(0)}{s_j(1)-s_j(0)+w_{i,j}(1)-w_{i,j}(0)}$$
These two equations clearly show that in control-based approach the disturbances enter the synchronization error dynamics linearly, while in the least-square dynamics non-linearly. One of the consequences is that, roughly speaking, under the first approach the global time synchronization error grows as the square root of the diameter $D$ of the WSN, i.e., $\max_{i,j}|e_{ij}(\infty)| \propto \sqrt{D}$, while under the second it grows exponentially when $H=2$, i.e.,
$\max_{i,j}|e_{ij}(\infty)| \propto a^D$ for some $a>1$. This fact will be evident when comparing FloodPISynch and FTSP in Section~\ref{sec:Experiments}. Also note that in the control-based approach the measurement errors are amplified by the control gains $\alpha_j,\beta_j$, which therefore should be kept small to reduce the steady state error. However, this comes at the price of slower convergence rate since the modulus of the largest eigenvalue of $F$ approaches unity for $\alpha_j\to 0,\beta_j\to 0$.

\item Memory and CPU requirements for the control-based strategy are minimal since only two additions and two multiplication are needed and no data storage is necessary. Differently, in the least-squares-based approach it is necessary to store $2H$ measurements and perform $3H$ multiplications, $3H$ additions and a $2\times 2$ matrix inversion and multiplication. This difference becomes even more pronounced in the fully-distributed scenario in WSN since, while in the control-based strategy memory and CPU requirements are the same, in the least-squares strategy each node has to store $2H$ measurements from each of its neighbours. If we indicate with $M$ the average number of neighbours per node, then memory and CPU complexity for the least-squares strategy is $O(MH)$ and $O(H)$, respectively, while it is only $O(1)$ for the control-based strategy. 

\item When considering general WSNs with a large number of nodes, Eqn.~\eqref{eqn:alpha_beta} might not be valid since they have been derived by considering only two nodes. Indeed, under an ideal scenario of synchronized communication in the WSN, the stability condition becomes
$$ 0<\beta_j<1, \ \  0<\alpha_j< \kappa(\mathcal{G})\frac{\beta_j}{\bar f_j B} $$
\end{itemize} 
where $\kappa(\mathcal{G})<1$ is a constant smaller than unity that depends on the topology of the WSN and can be arbitrarily small. As a consequence it is not possible to obtain ``universal'' stabilizing parameters that are guaranteed to work for any WSN. To overcome this problem, we will propose an \emph{adaptive Integral gain} that is disabled if the local synchronization errors are large, i.e., $\alpha_j(h)=0$ if $|e_{ji}(h)|>e_{max}$. Similarly as shown above, it is possible to show that if $\alpha_j=0$ and $0<\beta_j<2$, the synchronization error dynamics is stable, although a steady state error is present due to the uncompensated clock speeds. Nonetheless, this guarantees stability for any network topology.

\subsection{Design of the control parameters $\alpha$ and $\beta$}
\label{sec:parameter-adaptation}

In this section we provide an adaptive law for the real-time design of the parameter $\beta_j$ and $\alpha_j$, $j \in \left\{1,\ldots,N\right\}$, based only on the local information that node $j$ receives from its neighbours. 

We start with the \textbf{proportional gain} $\beta_j$. Loosely speaking, the role of input $u'_j$ in \eqref{eq:control-u} is that of reducing the distance between the logical clock $\hat{t}_i(T_{\mathrm{tx},i})$ and the logical clock $\hat{t}_j \left(T_{\mathrm{tx},i}+\gamma_{i,j}\right)$. Observe that the first equation of \eqref{eq:Update_i1} can be conveniently rewritten as
$$
\hat{t}_j (\,\left(T_{\mathrm{tx},i}+\gamma_{i,j}\right)^+\,) = (1-\beta_j)\,\,\hat{t}_j(T_{\mathrm{tx},i}+\gamma_{i,j}) +  \beta_j  \hat{t}_i(T_{\mathrm{tx},i})
$$
If 
$0< \beta_j \leq 1$, then $\hat{t}_j(\,\left(T_{\mathrm{tx},i}+\gamma_{i,j}\right)^+\,)$ turns out to be a convex combination of $\hat{t}_j(T_{\mathrm{tx},i}+\gamma_{i,j})$ and $\hat{t}_i(T_{\mathrm{tx},i})$ and it is easy to verify that
$$
|\hat{t}_j\left(\,(T_{\mathrm{tx},i}+\gamma_{i,j})^+\,\right) - \hat{t}_i(T_{\mathrm{tx},i})| \leq |\hat{t}_j(T_{\mathrm{tx},i}+\gamma_{i,j})  - \hat{t}_i(T_{\mathrm{tx},i})|
$$
Although the condition $0< \beta_j \leq 1$ is more restrictive than ~\eqref{eqn:alpha_beta}, this has the advantage to guarantee that the global time synchronization error cannot increase for any network topology, i.e.
$$
\max_{\ell,m}|\hat{t}_\ell\left(\,(T_{\mathrm{tx},i}+\gamma_{i,j})^+\,\right) - \hat{t}_m(T_{\mathrm{tx},i})| \leq \max_{\ell,m} |\hat{t}_\ell(T_{\mathrm{tx},i}+\gamma_{i,j})  - \hat{t}_m(T_{\mathrm{tx},i})|
$$
 and therefore cannot lead to instability. Basically the role of the parameter $\beta$ is to reduce the difference between the logical clock of the node transmitting information and the node receiving that information.
 In the next section, when introducing AvgPISynch, FloodPISync and PulsePISync protocols we will assume that $\beta_j$ is constant and such that $0< \beta_j \leq 1$, for all $j \in \left\{1,\ldots,N\right\}$, i.e.
 \begin{equation}\label{eqn:beta_global}
\beta_j=\beta, \ \ \ 0< \beta \leq 1.
\end{equation}

For what concerns the design of the \textbf{integrator gain} $\alpha_j$, an adaptive strategy to improve both convergence rate and stability is proposed. As discussed in the previous Section~\ref{subsec:control_vs_least_squares} the main role of the integrator is to compensate for the different clock speeds, whose absence would otherwise give rise to the typical saw-tooth behaviour of the time synchronization error. However, the integrator feedback comes at the price of few limitations. First, the rate of convergence is typically reduced as compared to a proportional feedback only. Second, if $\alpha_j$ is too large, it might drive the algorithm to instability. Third, integrator feedback suffers from the so called windup problem, i.e. the integrator is active even when the measured synchronization is not due to the different clock speeds, but only to large offset differences (see Section~3.5 in \cite{Astrom:95}). Consider for example, the same scenario as in the previous subsection with only two nodes where $\bar f_i<\bar f_j$ and $ e_{ji}(0)=\hat{t}_{j}(0)-\hat{t}_{i}(0)>0$, i.e. node $i$ is slower than node $j$, but the initial clock offsets are such that $\hat{t}_j(0)>\hat t_i(0)$. As a consequence, if node $i$ sends a message to node $j$, then update at node $j$ is such that $\hat \Delta_j(1) = \hat \Delta_j(0)+\alpha_j e_{ji} >  \hat \Delta_j(0)$, i.e. the speed of the logical clock $j$ is increased even further and initially the time synchronization error between these two nodes is likely to increase rather than decrease. Eventually, this situation is reversed and the PI controller will synchronize the two clocks, but the initial transient performance is degraded. To overcome this limitation, we propose to use an adaptive integrator gain $\alpha_j$, whose value is calculated by considering the derivative of the local synchronization error $e_{ji}$ and the derivative of the rate multiplier $\hat \Delta_{ji}$  according to the following equation:
\begin{equation} \label{eqn:new_adaptation_rule}
  \alpha_j(h) = 
  \begin{cases}
   	 0 & \mbox{if } |e_{ji}(h)|> e_{max} \\ 
	 \alpha_{max} & \mbox{if } |e_{ji}(h-1)| > e_{max} \mbox{ and }  |e_{ji}(h)| \leq e_{max}\\
	\lambda_j(h) \alpha_j(h-1) & \mbox{otherwise}
   \end{cases}    
\end{equation} where 

\begin{equation} \label{lambda}
	\lambda_j(h) = 
		\begin{cases}
    		1 & \mbox{if } e_{ji}(h-1)= 0 \mbox{ or }  e_{ji}(h-1) = e_{ji}(h-1)\\
	    	\min\left\{ \left|\dfrac{e_{ji}(h-1)}{e_{ji}(h)-e_{ji}(h-1)} \right|, 		\frac{\alpha_{max}}{\alpha_{j}(h-1)}\right\} & \mbox{otherwise}.
  \end{cases}    
\end{equation}

The intuition behind this approach can be explained as follows: As long as $e_{ji}(h)$ is greater than $e_{max}$, the integrator is disabled, since the observed error is mainly due to large initial offsets. This typically occurs at the on start of the synchronization. Whenever it is required to turn on the integrator since $e_{ji}(h)<e_{max}$, then $\alpha_j(h)$ is set to $\alpha_{max}$. As shown in Section~\ref{subsec:pairwise_control}, the pair $\beta=1, \alpha=\frac{1}{\hat f B}$ provides the faster rate of convergence. However, this choice also gives rise to the largest steady state error as shown in Section~\ref{sec:noise2}, therefore we would need a strategy that would gradually reduce the value of the gain $\alpha$ at steady state. The multiplicative parameter $\lambda(h)$ serves this objective. In fact, if the observed error $e_{ij}(h)$ is due to uncompensated different clock differences, which typically occurs during the transient or after a sudden change in frequency of one or more nodes, then we have $ e_{ij}(h)\approx e_{ij}(h-1)$ and $|e_{ij}(h-1)|> |e_{ij}(h-1)-e_{ij}(h)|$, therefore $\lambda_j(h)>1$, which implies that $\alpha_j(h)>\alpha_j(h-1)$, i.e. the integrator gain is increased to quickly compensate clock differences. After the algorithm has converged to steady state, then the error will typically appear as white noise, i.e. $ e_{ij}(h)$ oscillates around zero. This implies that typically $|e_{ij}(h-1)|\approx |e_{ij}(h)|$ and $|e_{ij}(h-1)|< |e_{ij}(h-1)-e_{ij}(h)|$, therefore $\lambda_j(h)<1$, which implies that $\alpha_j(h)<\alpha_j(h-1)$, i.e. the integrator gain is decreased to reduce steady-state error. The condition $\lambda_j(h)=1$ if $e_{ji}(h-1)= 0$ is necessary to avoid that $\alpha_j(h)$ remains equal to zero forever, while condition $\lambda_j(h)=1$ if $e_{ji}(h-1) = e_{ji}(h-1)$ to avoid division-by-zero numerical problems. Finally the saturation condition, provided by the $\min$ operator, is used to avoid to set $\alpha_j>\alpha_{max}$ which as shown in Section~\ref{subsec:pairwise_control}, would negatively impact the performance by increasing both the time-to-converge and the steady-state error.

The design problem is now shifted to designing the parameters $\beta,\alpha_{max}$ and $e_{max}$. For what concerns the first two parameters , we propose to use the optimal values in terms of rate of convergence given by Eqn.~\eqref{eqn:alpha_star}, i.e.
\begin{equation}\label{eqn:alpha_star_global}
\beta=1, \ \  \ \ \alpha_{max}=  \frac{1}{\hat f B}
\end{equation} which under the ideal scenario would provide convergence to the steady-state in a finite number of iterations. 

As for the second parameter $e_{max}$, we observed in the previous subsection, that any difference in the clock speeds will appear as a steady state error $e_{\Delta f}=\frac{\bar f_i-\bar f_j}{\hat f} B$ if the integrator feedback is not applied. This implies that if $e_{max}<e_{\Delta f}$, then the integrator will be never active and the steady state error will not decrease. Since potentially there could be two neighbouring nodes such that $|f_i-f_j|=2\Delta \! f_{max}$ by assumption, this implies that  $e_{max}$ must be bigger that the previous quantity. On the other hand, a large value for $e_{max}$ would lead to the integrator windup, and therefore longer convergence rate, therefore $e_{max}$ should be chosen close to its lower bound, i.e.
\begin{equation}\label{eqn:emax_lower_bound}
e_{max}=   \frac{2\Delta\!f_{max} }{\hat f}B.
\end{equation}

\begin{figure} 
\center

\includegraphics[scale=0.7]{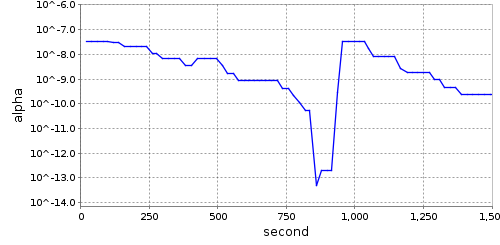} 

\caption{\label{fig:alpha_design} Plot of the evolution of $\alpha$ assuming $\beta_j=1$, $B=30$sec and $\hat f =1MHz$.}
\end{figure}

Figure \ref{fig:alpha_design} shows a numerical simulation using realistic values for the noise and topology. It can be seen that at the beginning $\alpha(h)$ is large but later is decreases, thus providing lower steady state error. However, at time 800, we manually changed the frequency of one clock, thus simulating a sudden change in frequency. As shown in the picture, the gain $\alpha$ rapidly increases and quickly compensate for it.

\section{Time Synchronization Protocols Based on PI Synchronization Algorithm}
\label{sec:Protocols}

In this section, we propose three time synchronization protocols for sensor networks based on the PISync algorithm. We first present a \textit{fully distributed} time synchronization protocol, in which each sensor node considers only the logical clock values of its neighbouring nodes to update the value and the rate of its logical clock. 
 Then, we present two \textit{flooding-based} time synchronization protocols in which each sensor node synchronizes to the clock of a reference node.

All of the protocols that we propose have very little memory requirement and are lightweight in terms of computation since they require only a few arithmetic operations to update the logical clocks, independently of network size and topology. Moreover, the amount of information to be exchanged among the sensor nodes is quite small since sensor nodes only broadcast the value of their logical clocks.

\subsection{Fully Distributed PISync Protocol: Average PISync Protocol (AvgPISync)}

In this subsection, we present a fully distributed PISync protocol, named Average PISync (\textit{AvgPISync}), which synchronizes each sensor node to their direct neighbours. The pseudo-code of this approach is presented in Algorithm \ref{alg:PI-dist}.

\begin{algorithm}

\label{alg:PI-dist}
1:~ $\square$ \textbf{Initialization} \\
2:~ $\hat{t}_{u}\leftarrow0$; $\hat{\Delta}_{u}\leftarrow 1/\hat{f}$; $sum\leftarrow0$; $num\leftarrow0$\\
3:~ \\
4:~ $\square$ \textbf{Upon receiving} $<\hat{t}_{v}>$ \\
5:~ $sum\leftarrow sum+(\hat{t}_{v} - \hat{t}_{u})$ \\
6:~ $num\leftarrow num+1$ \\
7:~\\
8:~ $\square$ \textbf{Upon} $s_{u}$ \textbf{is a multiple of $B\hat{f}$} \\
9:~  \textbf{if}  $|sum/num - \hat{t}_{u}| < e_{max}$ \textbf{then} $\hat{\Delta}_{u}\leftarrow \hat{\Delta}_{u} + \alpha_u(sum/num)$ \textbf{endif} \\
10:~$\hat{t}_{u}\leftarrow \hat{t}_{u} + (sum/num)$ \\
11:~$sum\leftarrow0$; $num\leftarrow0$ \\
12:~broadcast$<\hat{t}_{u}>$ 

\caption{AvgPISync pseudo-code for node $u$.}
\end{algorithm}

Each sensor node $u$ maintains two variables related to its logical clock: time estimate $\hat{t}_{u}$ and  oscillator frequency estimate $\hat{\Delta}_{u}$.\footnote{i.e. rate multiplier} Initially when the node is powered on, the time estimate is set to zero and the oscillator frequency estimate is set to $1/\hat{f}$, i.e. the nominal frequency, to progress the time estimate at the same speed of the hardware clock. Two other variables, $sum$ and $num$ are required to calculate the average synchronization error to the neighboring nodes. These variables are also initalized with zero (Lines 1-2).

Whenever a synchronization message from any neighbouring node is received (Line 4), the difference of the received time estimate $\hat{t}_{v}$ and the time estimate $\hat{t}_{u}$ of node $u$ at the reception time, i.e. \textit{clock skew},  is calculated. This value is added to the $sum$ variable (Line 5) and the number of received clock values is incremented (Line 6).

In order to inform its neighbouring nodes, node $u$ broadcasts its up-to-date time information approximately every $B$ seconds, where $B$ denotes the \textit{beacon period}. Whenever the hardware clock is a multiple of $B\hat{f}$ (Line 8), if the average clock skew $sum/num$ is greater than $e_{max}$, i.e. the maximum difference that can be observed due to different clock speeds between subsequent synchronization message reception, then the clock skew is mainly due to the large offsets between the clocks. Otherwise, the frequency of the oscillator is required to be adjusted (Line 9). Finally, the time estimate $\hat{t}_{u}$ of node $u$ is also updated (Line 10). It should be noted that the rate multiplier $\hat\Delta_u$ is updated by using the strategy given by  Eqn.~\eqref{eqn:new_adaptation_rule} and parameter $\beta$ is chosen to be equal to unity according to the Eqn. (\ref{eqn:alpha_star_global}). The variables $sum$ and $num$ are initialized (Line 11)  and the clock value is broadcasted (Line 12).

It should be noted that AvgPISync operates in a completely blind fashion since it requires neither to know the sender node nor to store its time information. Hence, AvgPISync is quite robust to topological changes and it can even work efficiently in topologies with very high densities.

\subsection{Flooding Based PISync Protocols: FloodPISync and PulsePISync}

We now present a flooding based version of the PISync protocol, namely Flooding PISync Protocol (\textit{FloodPISync}),  which synchronizes each sensor node to the clock of a reference node. In FloodPISync, a dynamically elected or a predefined reference node floods its stable time information into the network. Each sensor node collects this time information, updates its clock according to PISync algorithm and also broadcasts its clock value for its neighbouring nodes to achieve network-wide time synchronization. The pseudo-code of FloodPISync is presented in Algorithm \ref{alg:PI}. For simplicity, assume that $ref$ is the predefined reference node. 

\begin{algorithm}

\label{alg:PI}
1: $\square$ \textbf{Initialization} \\
2: $\hat{t}_{u}\leftarrow0$; $\hat{\Delta}_{u}\leftarrow 1/\hat{f}$; $seq_{u}\leftarrow0$  \\
3: \\
4: $\square$ \textbf{Upon receiving} $<\hat{t}_{v},seq_{v}>$ \textbf{such that} $seq_{u}<seq_{v}$ \\
5: \textbf{if} $|\hat{t}_{v} - \hat{t}_{u}| < e_{max}$ \textbf{then} update $\hat{\Delta}_{u}\leftarrow \hat{\Delta}_{u} + \alpha_u(\hat{t}_{v} - \hat{t}_{u})$ \textbf{endif}\\
6:   $\hat{t}_{u}\leftarrow \hat{t}_{v}$ \\
7:   $seq_{u}\leftarrow seq_{v}$ \\ 
8: \\
9: $\square$ \textbf{Upon} $s_{u}$ \textbf{is a multiple of $B\hat{f}$} \\
10: \textbf{if} $u=ref$ \textbf{then} $seq_{u}\leftarrow seq_{u}+1$ \textbf{endif} \\
11: broadcast$<\hat{t}_{u},seq_{u}>$ 

\caption{FloodPISync pseudo-code for node $u$ with a fixed reference node $ref$.}
\end{algorithm}

Apart from the variables of the fully distributed version, each sensor node $u$ also maintains a sequence number $seq_{u}$ to store the largest sequence number received from the reference node. Initially when the node is powered on, the sequence number is also set to zero (Lines 1-2).

The reception of a synchronization message carrying a greater sequence number than $seq_{u}$ indicates that the reference node has initiated a new synchronization round recently (Line 4). Hence, the received time information can be considered as a fresh estimate of the reference clock. Similar to the fully distributed version, the difference of the received time estimate $\hat{t}_{v}$ and the time estimate $\hat{t}_{u}$ of node $u$ at the reception time is considered to update the oscillator frequency estimate $\hat{\Delta}_{u}$. Similar to AvgPISync, this update is performed by setting the parameter $\beta$ equal to unity. As a consequence, the neighbours whose time information have been received later have weight 0 and they are not considered to update time estimate. The time estimate $\hat{t}_{u}$ of node $u$ and the sequence number are also updated (Lines 6-7). 

Approximately every $B$ seconds, solely the reference node increments its sequence number and hence initiates a new flooding round. Since all of the sensor nodes broadcast their time estimates, time information of the reference node is propagated and network-wide synchronization is achieved (Lines 9-11). 

As can be observed, the  time information of the reference node may follow many paths to reach any node $u$ in any synchronization round. Node $u$ takes into account only the firstly received message since it can be considered as carrying the most up-to-date estimate of the reference clock. The other messages belonging to the same synchronization round are discarded by node $u$. In practice, this gives rise to a tree-like synchronization topology since each node tends to use data received by one its neighbouring nodes closer to the root.

In FloodPISync, each node does not propagate the time information of the reference node as soon as it receives it. Instead, nodes wait $s_{u}$ to reach a multiple of $B\hat{f}$. It has been shown in \cite{Lenzen2009Optimal} that waiting times at each hop increase the estimation errors. Moreover, in Section \ref{subsec:control_vs_least_squares}, we have shown that with least-squares the effect of various error sources appear in the time synchronization error dynamics as multiplicative noise. Hence, it is desirable to propagate fresh time information as fast as possible through the pulses. Based on this observation, we also propose a fast flooding based synchronization protocol as in \cite{Lenzen2009Optimal}, namely \textit{PulsePISync}, for architectures such as Glossy \cite{ferrari2011efficient} which allow efficient and reliable fast-flooding. The pseudo-code of PulsePISync which is quite similar to FloodPISync is presented in Algorithm \ref{alg:PulsePI}.

\begin{algorithm}

\label{alg:PulsePI}
1: $\square$ \textbf{Initialization} \\
2: $\hat{t}_{u}\leftarrow0$; $\hat{\Delta}_{u}\leftarrow 1/\hat{f}$; $seq_{u}\leftarrow0$  \\
3: \\
4: $\square$ \textbf{Upon receiving} $<\hat{t}_{v},seq_{v}>$ \textbf{such that} $seq_{u}<seq_{v}$ \\
5: \textbf{if} $|\hat{t}_{v} - \hat{t}_{u}| < e_{max}$ \textbf{then} update $\hat{\Delta}_{u}\leftarrow \hat{\Delta}_{u} +  \alpha_u(\hat{t}_{v} - \hat{t}_{u})$ \textbf{endif}\\
6:   $\hat{t}_{u}\leftarrow \hat{t}_{v}$ \\
7:   $seq_{u}\leftarrow seq_{v}$ \\ 
8:	 broadcast$<\hat{t}_{u},seq_{u}>$ \\
9: \\
10: $\square$ \textbf{Upon} $s_{ref}$ \textbf{is a multiple of $B\hat{f}$} \textbf{at the reference node $ref$}\\
11: $seq_{ref}\leftarrow seq_{ref}+1$ \\
12: broadcast$<s_{ref},seq_{ref}>$ 
\caption{PulsePISync pseudo-code for node $u$ with a fixed reference node $ref$.}
\end{algorithm}

The main difference of PulsePISync from FloodPISync is that each node propagates its current time information as soon as it receives the time information of the reference node (Line 8). Hence, the propagated time information suffers only from the various error sources through communication, but not from the estimation errors. Only the reference node $ref$  broadcasts the value of its hardware clock together with its current sequence number every $B$ seconds (Lines 10-12). Hence, each node waits an up-to-date time information from the reference node and broadcasts that information as fast as possible to reduce the estimation errors due to the waiting times.

\begin{figure} 
\center

\includegraphics[scale=0.5]{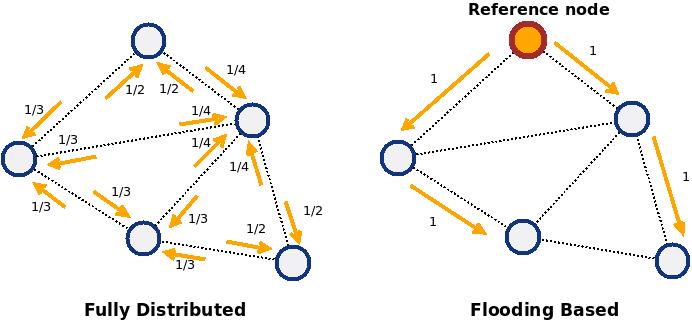}

\caption{\label{fig:comparison} Comparison of the PI synchronization protocols in terms of the weights given to the neighbours during time information update.}
\end{figure}

A comparison of the fully distributed AvgPISync and flooding based FloodPISync and PulsePISync in terms of weights given to the neighbours during time information update is presented in Figure \ref{fig:comparison}. In AvgPISync, time information received from neighbouring nodes have equal weights when updating current time information, since the average synchronization error is considered. On the other hand, in flooding based protocols FloodPISync and PulsePISync, only one of the neighbouring nodes which propagated current time information of the reference node earliest during the current synchronization round is considered for this update, namely the node from which the first largest sequence number is received. All other nodes are neglected. 

\subsection{Analysis in presence of transmission noise for flooding-based PISync protocols}
\label{sec:noise}

In this section we consider the Flooding based PISync Protocol in presence of transmission noise and time-varying frequencies. We restrict our analysis to a line graph (see Figure \ref{fig:topology}) with $N$ nodes, labelled $1$ through $N$. We assume that node $i+1$ receives information from node $i$ and that node $1$ is the reference node.

We denote the different sources of noises with the same formalism adopted in Section \ref{sec:noise2}. In particular we assume that noises related to different transmissions are independent from each other, but have the same variance. 

Let us consider node $2$. Similarly to what done in Section \ref{sec:noise2}, let $e_2(h)=\hat{t}_2(h)-\hat{t}_1(h)$ and let $P_{e_2}(h)=\mathbb{E} \left[e^2_2(h)\right]$. The asymptotic value of the mean-square error of $e_2$, i.e., $\bar{P}_{e_2}=\lim_{h \to \infty} P_{e_2}(h)$, is given in \eqref{eq:AsympError}.

Now let node $3$ be a child of node $2$. We have that a time $hB$ node $3$ receives from node $2$ the information $\hat{t}_1(h)+v_{1,2}(h)+v_{2,3}(h)$. Let us define $e_{3}(h)=\hat{t}_3(h)- \hat{t}_1(h)$ and, accordingly, $P_{e_{3}}(h)= \mathbb{E}\left[e_{3}^2(h)\right]$. Since 
noises related to different transmissions are independent from each other and since all the noises are assumed to have the same variance, we obtain that
$$
\bar{P}_{e_3}=\lim_{h \to \infty} P_{e_3}(h)=2\bar{P}_{e_2}.
$$
The above reasoning can be extended to any node in the graph. Specifically, denoting by $\bar{P}_{e_i}$ the asymptotic variance of the synchronization error between node $1$ and node $i$, we have that
$$
\bar{P}_{e_i}=d\bar{P}_{e_2},
$$
where $d$ is the distance on the graph between node $1$ and node $i$. In other words the synchronization error variance (resp., the standard deviation) grows linearly (resp., as the square root) with the distance to the reference node.
Some remarks are now in order:

\begin{itemize}
\item Observe that, from Eqn. \eqref{eq:AsympError}, it follows that, the smaller the value of $\alpha$ is, the smaller the value of the asymptotic synchronization error is. Nonetheless it is worth recalling that, as $\alpha$ approaches $0$, the algorithm becomes slower and slower to converge. 

\item In the previous analysis, we considered only transmission errors and time-varying frequencies. 
However, our analysis could be extended to include also the presence of transmission delays, in the case an estimate of  the mean of the delays is available and a correcting term trying to compensate the effects due to the delivery delay is applied in the control law. More precisely, observe that
$$
\hat{t}_j(T_{\mathrm{tx},i})= \hat{t}_j(T_{\mathrm{tx},i}+\gamma_{i,j})- \gamma_{i,j} \bar{f}_j\hat{\Delta}_j(T_{\mathrm{tx},i})
$$
where the unknown quantities $\gamma_{i,j}$ and $\bar{f}_j$ might be suitably approximated by $\bar{\gamma}$ and $\hat{f}$, being $\bar{\gamma}$ a known estimate of the mean of the delay $\gamma_{i,j}$\footnote{The information about $\bar{\gamma}$
 might be available from a-priori statistical studies as suggested in \cite{ZZ-PPC-TH:09}.}. In this way we would obtain
$$
\hat{t}_j(T_{\mathrm{tx},i}) \approx \hat{t}_j(T_{\mathrm{tx},i}+\gamma_{i,j})- \bar{\gamma} \hat{f}\hat{\Delta}_j(T_{\mathrm{tx},i})
$$
and the control inputs in \eqref{eq:control-u} might be modified as
\begin{align}\label{eq:control-u-modified}
u'_j &=   \beta_j \cdot \left(\, \hat{t}_i(T_{\mathrm{tx},i})-\hat{t}_j \left(T_{\mathrm{tx},i}+\gamma_{i,j}\right)+ \bar{\gamma} \hat{f}\hat{\Delta}_j(T_{\mathrm{tx},i})\,\right)\nonumber\\
u''_j &=   \alpha_j \cdot \left(\, \hat{t}_i(T_{\mathrm{tx},i})-\hat{t}_j \left(T_{\mathrm{tx},i}+\gamma_{i,j}\right)+ \bar{\gamma} \hat{f}\hat{\Delta}_j(T_{\mathrm{tx},i})\,\right)
\end{align}
Observe that, in the above model, the transmission delays can be treated as noises of zero-mean.
\item  A similar analysis might be provided also for the AvgPISync algorithm following Section V of \cite{Carli_2011}. However in \cite{Carli_2011}, few unrealistic conditions are assumed, i.e., all the transmitting and updating steps are performed synchronously and all the clocks have the same frequency. To generalize this analysis to the AvgPISync algorithm is quite complicated and goes beyond the scope of this paper. We limit ourselves to the numerical results reported in Section \ref{subsec:AvgPISync-GTSP}.
\end{itemize}

\section{Real-world Experiments and Simulations}
\label{sec:Experiments}

In this section, we present an experimental evaluation of AvgPISync, FloodPISync and PulsePISync in our testbed of 20 sensor nodes. We first present the experimental results of the fully distributed protocol AvgPISync and its comparison with another fully distributed protocol GTSP. Then, we present experimental results related to the flooding based protocols FloodPISync and PulsePISync, and their comparison with other flooding based time synchronization protocols FTSP and PulseSync. Finally, we present a comparison of these protocols through simulations in terms of their scalability by considering networks with larger diameters.

\subsection{Hardware Platform and Implementation Details}

The experimental platform of our experiments is based on a WSN composed by MICAz nodes from Memsic\footnote{http://www.memsic.com/} instrumented with a 7.37Mhz 8-bit Atmel Atmega128L microcontroller. MICAz nodes are equipped with 4kB RAM, 128kB program flash and Chipcon CC2420 radio chip which provides a 250 kbps data rate at 2.4 GHz frequency. We used 7.37 MHz quartz oscillator on the MICAz board as the clock source for the timer used for timing measurements. The timer operates at 1/8 of that frequency and thus each timer tick occurs at approximately every 921 kHz, i.e. approximately 1 microsecond.

For our real-world experiments, we used the publicly available implementation of FTSP coming with TinyOS 2.1.2.\footnote{http://www.tinyos.net} We also implemented PulseSync, GTSP, AvgPISync, FloodPISync and PulsePISync ourselves in TinyOS. The CC2420 transceiver on the MICAz board has the capability to timestamp synchronization packets at MAC layer  with the timer used for timing measurements. This is a well-known method that increases the quality of time synchronization by reducing the effect of the non-deterministic error sources arising from communication \cite{Maroti2004}. We used the packet level time synchronization interfaces provided by TinyOS to timestamp synchronization messages at MAC layer \cite{Packet-level-time-synchronization:2008}.

\subsection{Testbed Setup}

In our testbed, sensor nodes are placed in the communication range of a \emph{reference broadcaster}. This node periodically broadcasts \emph{query packets} which will be received approximately at the same time by all nodes. Each sensor node transmits a \emph{reply packet} carrying its logical clock value at the reception time of the query packet. A \emph{base station} node connected to the PC collects these reply packets and forwards them to the serial port for logging. At the end of the experiments, we analyze the logged experimental data. 

\begin{figure}
\center

\includegraphics[scale=0.4]{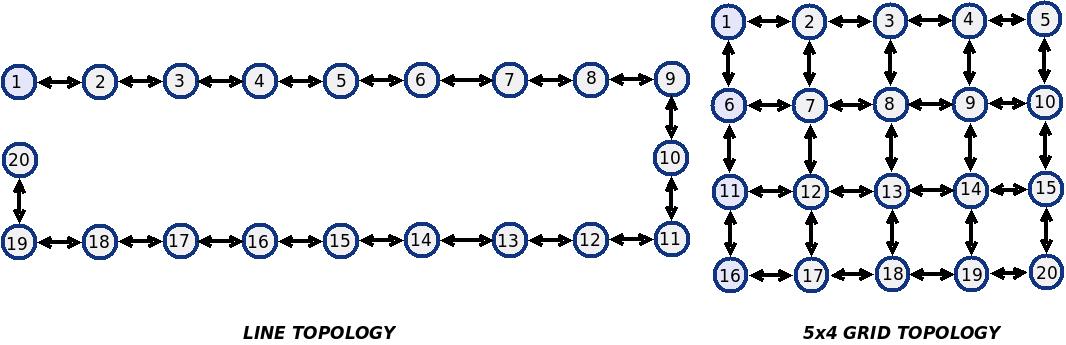}

\caption{\label{fig:topology} The placement of sensor nodes in order to construct the line
and grid topologies for the experiments.}
\end{figure} 

We constructed line and grid topologies by forcing nodes to accept only the packets from their neighbouring nodes in the corresponding topology. Sensor nodes discard the packets from the other nodes. The grid topology has been selected to evaluate performance of all protocols in moderately dense WSN. In this scenario, fully-distributed time synchronization approaches are likely to perform better than flooding-based synchronization approaches since contention, congestion and packet collisions are more likely to occur in this topology. Moreover, experimental results of the line topology are also important to observe the scalability of both fully distributed and flooding based schemes, since it has been shown in \cite{Lenzen2009Optimal} that the synchronization error grows with the network diameter. The placement of sensor nodes during the experiments is shown in Figure \ref{fig:topology}.

\subsection{Experimental Parameters}

During our experiments, the beacon period $B$ was 30 seconds for all of the protocols. For FTSP, GTSP and PulseSync, the capacity of the least-squares regression table was fixed to $H=8$ elements. Each sensor node was programmed by assigning an unique ID between 1 and 20. Since it takes several rounds for FTSP to dynamically elect the reference node, we modified it to work with a fixed reference node. For the flooding based protocols PulseSync, FTSP, FloodPISync and PulsePISync, the node with ID "1" has been set as reference node for the other nodes in the network. At the beginning of the experiments, sensor nodes are powered at random times within the first 2 minutes of operation and each experiment took 10000 seconds (approximately 2.8 hours).

The adaptation of the $\alpha$ parameter is a major point to consider in the practical implementation of the PISync algorithm. We followed the strategy presented in Section \ref{sec:parameter-adaptation} to adjust its value. Since the reported drift of the sensor nodes is defined $\pm 100$ ppm, i.e. $\Delta f_{max}/\hat f = 100\cdot 10^{-6}$ and the communication period $B$ is 30 seconds, then according to  Eqn.~\eqref{eqn:emax_lower_bound} we should have $$e_{max} = 0.006 sec \ (\approx6000 ticks)$$ and according to Eqn.~\eqref{eqn:alpha_star_global} we should have
$$\beta=1, \ \  \ \  \alpha_{max} = 3,33\cdot 10^{-8}.$$
Table \ref{tab:PI-Experimental-Parameters} summarizes the parameter values used for the PI synchronization protocols during our experiments.

\begin{table}

\center
\caption{Experimental Parameters used for the PI synchronization protocols.\label{tab:PI-Experimental-Parameters}}

\begin{tabular}{cccc}
$\bf{\beta}$ & $\bf{B}$ & $\bf{e_{max}}$ & $\bf{\alpha_{max}}$  \\ 
1 & 30 seconds & 0.006 seconds & $3,33\cdot 10^{-8}$ 
\end{tabular}

\end{table}

\subsection{Synchronization Performance metrics}

In order to evaluate the synchronization performance, we consider local and global synchronization errors. More precisely, we define \textit{instantaneous global skew of a node} as the maximum difference of the logical clock values observed between that node and any other arbitrary node at a given time instant. Similarly, we define \textit{instantaneous local skew of a node} as the maximum difference of the logical clocks observed between that node and any of its neighbouring nodes at a given time instant. 

By considering instantaneous global and local skews of nodes, we used the following metrics for the evaluation of the aforementioned protocols: 
\begin{itemize}

\item \textit{Instantaneous maximum global skew of the network} is defined as the maximum global skew among the nodes at a given time instant, i.e. $$MGS(t):=\underset{i,j \in V}{max}\lbrace\hat{t}_i(t)-\hat{t}_j(t)\rbrace$$ 

\item \textit{Instantaneous average global skew of the network} is defined as the average of the global skews of the nodes at a given time instant, i.e. $$AGS(t)=\frac{1}{N}\underset{i \in V}{\sum}\underset{j \in V}{max}\lbrace\hat{t}_i(t)-\hat{t}_j(t)\rbrace$$

\item 
\textit{Instantaneous maximum local skew of the network} is defined as the maximum local skew among the nodes at a given time instant i.e. $$MLS(t):=\underset{(i,j) \in \mathcal{E}}{max}\lbrace\hat{t}_i(t)-\hat{t}_j(t)\rbrace$$

\item 
\textit{Instantaneous average local skew of the network} is defined as the average of the local skews of the nodes at a given time instant, i.e. 
$$ALS(t):=\frac{1}{N}\underset{i \in V}{\sum}\underset{(i,j) \in \mathcal{E}}{max}\lbrace\hat{t}_i(t)-\hat{t}_j(t)\rbrace$$
\end{itemize}

The first two metrics $MGS$ and $AGS$ are important to understand how tight is the synchronization across the whole network. Typically, flooding-based synchronization methods perform well in these metrics since rapidly propagate information across the network. The second two metrics $MLS$ and $ALS$ are important to evaluate how tight is the synchronization among neighbours. These metrics are particularly important in applications where local synchronization rather than global synchronization is critical such as TDMA-based communication protocols. In general, fully distributed synchronization  methods perform better since they use all packets from the neighbours to set their logical clocks. Differently, in flooding-based methods only one packet is used, therefore physically closed neighbours can have larger synchronization errors than in fully distributed methods (see Fig.~\ref{fig:comparison}).

\subsection{ Effect on performance of  $\alpha_{max}$ and $e_{max}$ parameters}
\label{sec:eval-alpha}
Before presenting a detailed analysis of the aforementioned protocols in terms of their energy consumption, memory overhead and synchronization performance, we first evaluate the effects of the parameter $\alpha_{max}$  and $e_{max}$ defined in Section \ref{sec:parameter-adaptation} in terms of rate of convergence and steady state synchronisation errors based on real-world experiments. Based on the theoretical analysis in Section~\ref{subsec:control_vs_least_squares} and Section~\ref{sec:noise}, the smaller  $\alpha_{max}$, the smaller the global skew (\textit{steady state error}) but the longer the time took to achieve network-wide tight synchronization (\textit{time to convergence}). In order to show this effect in practice, we performed experiments with the FloodPISync on the line topology of 20 sensor nodes with different  $\alpha_{max}$ parameters. Table \ref{tab:alpha} presents steady state error and time to convergence values observed during these experiments. 

\begin{table}
\center
\caption{Steady state error and Time to convergence for different $\alpha_{max}$ values on a line topology of 20 MICAz sensor nodes executing FloodPISync.\label{tab:alpha}.}

\begin{tabular}{cccc}
$\bm{\alpha_{max}}$ & $\bm{e_{max}}$ & \textbf{Steady state error (AGS)} & \textbf{Time to convergence} \\
  $\bf{3,33 \cdot 10^{-9}}$ & 0.006 sec &  5 $\mu$sec & $\simeq$ 2000 sec\\ 
  $\bf{3,33 \cdot 10^{-8}}$ & 0.006 sec &  6 $\mu$sec & < 500 sec \\ 
  $\bf{3,33 \cdot 10^{-7}}$ & 0.006 sec &  > 10000 ticks & no convergence \\ 
\end{tabular}

\end{table}

\begin{figure}
\center

\includegraphics[scale=0.5]{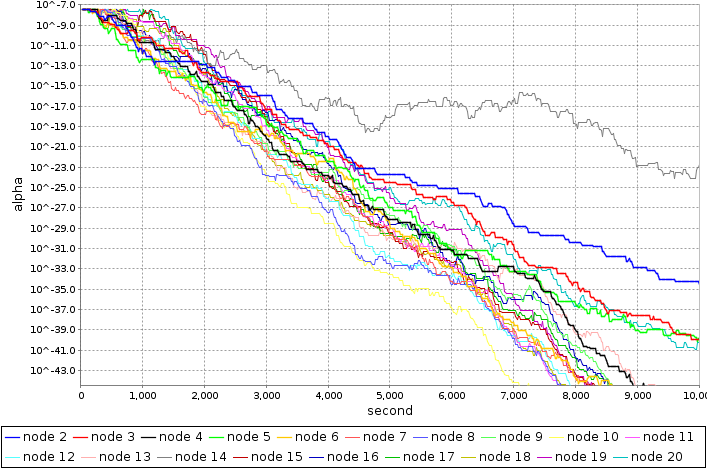}

\caption{\label{fig:alpha-evolution} Evolution of the $\alpha$ parameter on the $5\times 4$ grid topology of 20 MICAz sensor nodes for FloodPISync.}
\end{figure}

Figure \ref{fig:alpha-evolution} illustrates the evolution of the $\alpha$ parameters  according to the adaptive strategy given by Eqn.~\eqref{eqn:new_adaptation_rule} when  $\alpha_{max} = 3.33\cdot10^{-8}$. Initially, there are big differences between the logical clocks of the sensor nodes and this difference is  mainly due to the large offsets.  Hence, the $\alpha$ parameters of the nodes are zero and the integrator part of the PISync algorithm is off. When the integrator is turned on
and the offset differences tends to get smaller, the $\alpha$ values decrease from $\alpha_{max}$ to very small values close to zero.

As can be observed, with the maximum value of the $\alpha_{max}$ we presented in Table \ref{tab:PI-Experimental-Parameters} and the adaptive strategy given by Eqn.~\eqref{eqn:new_adaptation_rule}, we get tight synchronization with a reasonable time to convergence and hence these values are quite satisfactory for the practical implementation.

Another observation, based on experiments not reported here in the interest of space, is that choosing $e_{max}$ to be smaller than the lower bound given by inequality of Eqn.~\eqref{eqn:emax_lower_bound} would lead to nodes that do not turn on their integrator so that different clock speeds are not compensated for, thus leading to poor steady state synchronisation performance. Conversely, a large value of $e_{max}$ leads to a longer convergence late since integrator is turned on too early and exhibits the so called wind-up behaviour. Hence, setting $e_{max}$ to its lower bound value as presented in inequality \ref{eqn:emax_lower_bound} is quite satisfactory in practical implementations.

In conclusion, the selection of the design parameters $\alpha_{max}$ and $e_{max}$ and the adaptive gain strategy proposed in Section \ref{sec:parameter-adaptation} are quite satisfactory for the real-world implementation of the PISync protocols, as also confirmed by the experimental results in the following subsections.

\subsection{Evaluation of Energy Consumption and Memory Requirements}

We first start to evaluate PISync protocols, GTSP, FTSP and PulseSync by considering their energy consumption and memory requirements. From practical perspective, computation, communication and memory overheads of a time synchronization protocol are important points to consider. For flooding based time synchronization protocols, the amount of memory which is allocated to store collected reference time information determines their major main memory requirements (RAM). For fully distributed protocols which require to keep track of the neighbouring nodes, the amount of memory which is allocated to store neighbourhood information determines their major main memory requirements. Since the flash memory (ROM) is also a scarce resource which stores the program code of the applications, the overhead of the time synchronization protocol on the application code size must also be taken into account. An important point to consider in terms of energy consumption is the communication frequency and the length of the synchronization messages: in fact more frequent communication and longer synchronization messages give rise to higher energy consumption to transmit and receive these messages.  Moreover, if the length of the synchronization messages are short, i.e. the amount of information to be exchanged to achieve synchronization is small, then this enables to piggyback synchronization data to messages belonging to the application and other protocols. Computation overhead is another factor which affects the energy consumption: from the reception time of the reference time information to the time it is processed and logical clock is updated, considerable amount of energy is required. Table \ref{tab:Resource} summarizes the CPU overhead, synchronization message length and memory requirements of FTSP, PulseSync, GTSP and PISync protocols. 

\begin{table}
\center

\caption{CPU overhead, synchronization message length and memory requirements of FTSP, PulseSync, GTSP and PISync protocols during the experiments. $|\mathcal{N}|$ is used to denote maximum neighbourhood cardinality. \label{tab:Resource}}

\begin{tabular}{lcccc}
\multicolumn{1}{l}{} & \textbf{FTSP} & \textbf{PulseSync} & \textbf{GTSP} & \textbf{PISync} \\
\textbf{CPU Overhead }(ticks) & {\scriptsize $\simeq$ 5440} & {\scriptsize $\simeq$ 5440} & {\scriptsize $\simeq$ 5610 } & {\bf \scriptsize $\simeq$ 145 }\\
\textbf{Message Length} (bytes)& {\scriptsize 9  } & {\scriptsize 9  } & {\scriptsize 14 } & {\bf \scriptsize 4-9 }\\ 
\textbf{Main Memory Overhead} (bytes) & {\scriptsize 52  } & {\scriptsize 52 } & {\scriptsize 64{*}$|\mathcal{N}|$ + 12} & {\bf \scriptsize 16}  \\
\textbf{Flash Memory Requirements} (bytes)& {\scriptsize {18000} } & {\scriptsize {17856} } & {\scriptsize {22092} } & \bf {\scriptsize {15432 }}\\ 
\end{tabular}

\end{table}

\begin{table}
\center

\caption{The current consumption of the MICAz platform. The numeric values are taken from Figure 6 in
\cite{Telos:2005}. \label{tab:The-current-consumptions}}

{
\begin{tabular}{|l|r|}
\hline 
\textbf{Operation} & \textbf{MICAz}\\
\hline 
Minimum Voltage  & 2.7 V\\
\hline 
Standby & 27.0 $\mu$A\\
MCU Idle & 3.2 mA\\
MCU Active & 8.0 mA\\
MCU + Radio RX  & 23.3 mA\\
MCU + Radio TX (0 dbm) & 21.0 mA\\
\hline
\end{tabular}
}
\end{table}

Upon receiving a new synchronization message from the reference node, FTSP and PulseSync are required to store recently received time information into a regression table of 8 elements and perform a least-squares regression. This calculation consists of many integer multiplication and floating point division operations. GTSP not only performs these steps, but also it calculates the average of the clock offset and rate multipliers of the neighbouring nodes by considering stored neighbourhood time information. These operations consist of additional arithmetic operations on floating point numbers. For PISync protocols, only a few arithmetic operations are required and hence the CPU overhead of time synchronization decreases significantly. PISync decreases the computation overhead by more than 97\% compared to FTSP, PulseSync and  GTSP. The computation overhead also leads to overhead in terms of energy-efficient duty-cycling operation of the WSNs. Consider Table \ref{tab:The-current-consumptions} which presents the current consumptions in MICAz platform. When the micro-controller is active, the power consumption is approximately 21.6 mW, on the other hand, while receiving, the power consumption is approximately 62.91 mW. When Table \ref{tab:Resource} is considered,  processing a synchronization message will consume more or less 3132 nano Joules for PISync, while it will consume more or less 117504 nano Joules for PulseSync and 146880 nano Joules for GTSP. If it takes approximately 1500 microseconds to receive a TinyOS packet\footnote{In TinyOS, packets are composed of 11 bytes header, 28 bytes payload and 7 bytes footer, a total of 46 bytes. With 250 kps data rate, it takes approximately 1500 $\mu$ secs to send and receive these packets.}, the power consumption while receiving will be approximately 94365 nano Joules. Hence, in PulseSync and GTSP, the energy consumed for processing any synchronization message is greater than that consumed for transmitting or receiving any packet. This shows that CPU overhead is also an important factor to be considered in the design of time synchronization protocols. PISync protocols are lightweight in terms of energy consumption considerably.

\begin{figure}

\center

\includegraphics[scale=0.92]{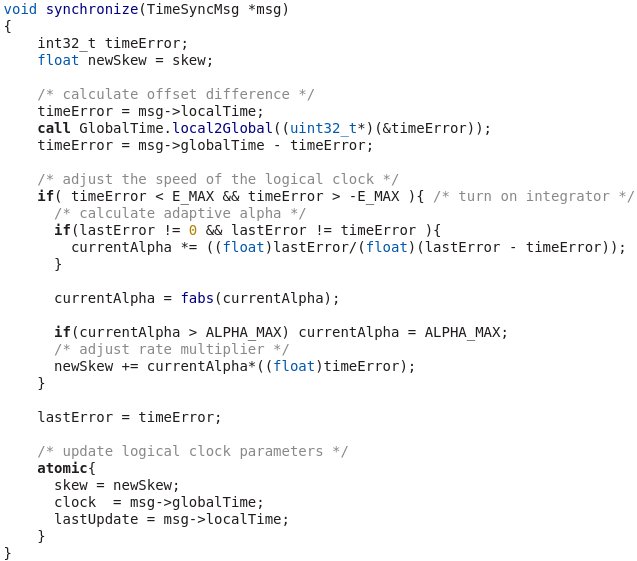}

\caption{\label{fig:tinyos} The main lines of the TinyOS implementation of FloodPISync protocol where the received time information is processed and logical clock is updated. The \textit{localTime} field of the message, which stores the hardware clock value of the receiver node when this message was sent, is used to calculate the logical clock value at that time via the \textit{local2Global} interface. To calculate the clock skew, this value is subtracted from the \textit{globalTime} field of the message which holds the sender's logical clock value. If the skew is smaller than $e_{max}$ value, the integrator gain is turned on and the adaptive value of the $\alpha$ parameter is calculated. Finally, the offset compensation is performed via the proportional gain and the logical clock parameters are updated.}
\end{figure}

For FTSP and PulseSync, synchronization messages carry the identifiers of the reference node and the sender node, sequence number and the logical clock value of the current node. A total of 9 bytes are required to carry this information. For flooding based versions FloodPISync and PulsePISync, the carried information is identical to that in FTSP and PulseSync. Hence, the length of synchronization messages is 9 bytes. For fully distributed version AvgPISync, there is no need to carry reference node and current node identifier, and the sequence number, since there is not any reference node and any flooding round. Each node just processes any message received from any of its neighbouring nodes. Hence, the synchronization messages are quite short in length, i.e. only 4 bytes. If GTSP is considered, since each node keeps track of its neighbouring nodes and stores their time information, sender node identifier, its logical clock value, hardware clock value and the value of the rate multiplier must be included in the synchronization messages. Hence, the length of synchronization messages is 14 bytes in GTSP. It can be concluded that AvgPISync decreases the amount of carried information approximately by 70\%  when compared to another fully distributed protocol GTSP.

Considering Eqn. (\ref{eq:hat_t}), the aforomentioned protocols require to store the value of their hardware clock $s(t_0)$, the updated value of their logical clock $\hat{t}(t_0)$ and their rate multiplier $\hat{\Delta}(t_0)$ at the latest update time $t_0$. Each of these variables are 4 bytes, and hence 12 bytes of main memory is required to the variables related to the logical clock. As mentioned previously, FTSP and PulseSync store the time information of the reference node at a regression table of 8 pairs. In practical implementation, 40 bytes of memory is allocated for this table. For GTSP, 64 bytes of memory is allocated to keep track each neighbouring node, which is an extremely big memory overhead. For PISync protocols, no memory storage is required to keep track of reference node in FloodPISync and PulsePISync,  and neighbouring nodes in the fully distributed AvgPISync. According to Eqn. (\ref{eqn:new_adaptation_rule}) which presents the adaptation of the alpha parameter, the error observed at the reception time of the last synchronization message, i.e. $e_{ji}(h-1)$, should be stored, which is 4 bytes. Hence, the total main memory overhead of PISync protocols are very small compared to the other protocols.

Finally, as shown in Table \ref{tab:Resource}, PISync protocols have significantly reduced application code size. Their implementations are really simple and only require quite a few arithmetic operations as shown in Figure \ref{fig:tinyos}. Since flash memory (ROM) which stores the program code of the applications is a scarce resource, PISync protocols led to a remarkable gain for time synchronization in terms of ROM requirements.

\subsection{ Performance comparisons: AvgPISync vs GTSP}\label{subsec:AvgPISync-GTSP}

\begin{figure*}

\center

\subfigure[Global Skew]{
\includegraphics[scale=0.5]{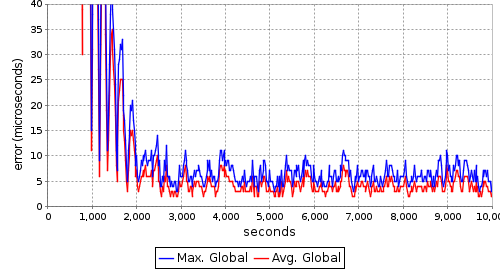}$\,
$\includegraphics[scale=0.5]{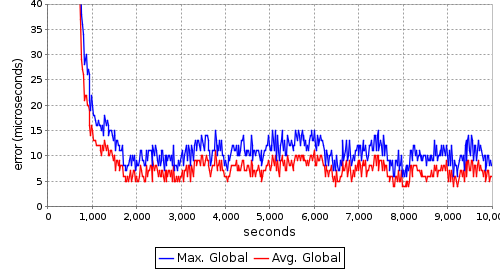}
}

\subfigure[Local Skew]{
\includegraphics[scale=0.5]{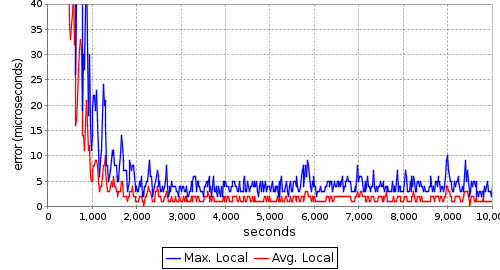}$\,
$\includegraphics[scale=0.5]{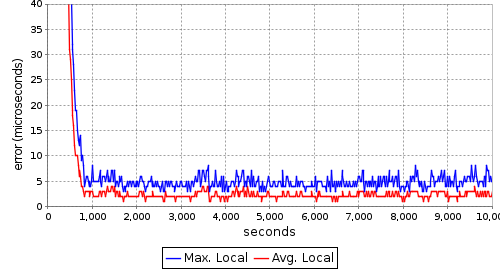}
}

\subfigure[Rate multipliers]{
\includegraphics[scale=0.5]{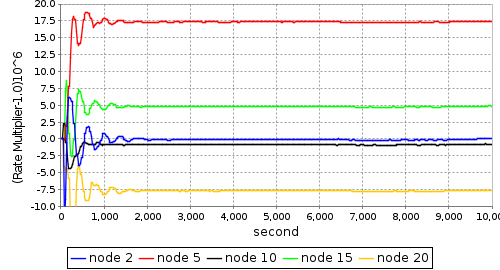}$\,
$\includegraphics[scale=0.5]{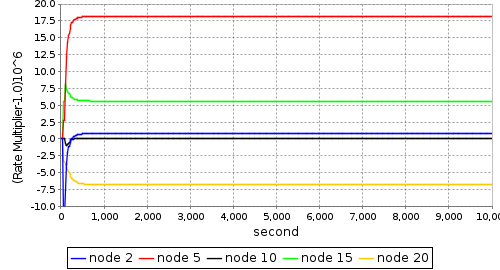}
}

\subfigure[Maximum local skew per node]{
\includegraphics[scale=0.5]{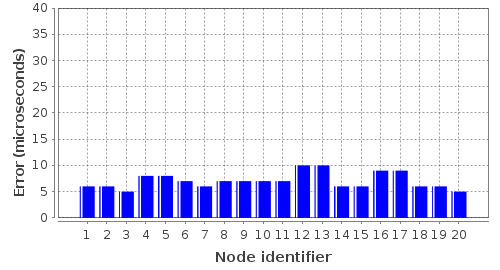}$\,
$\includegraphics[scale=0.5]{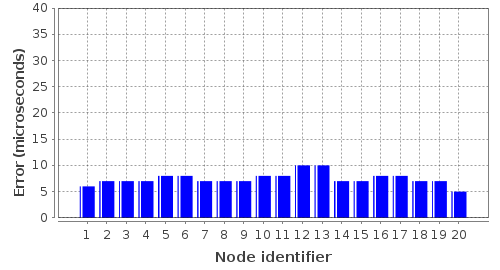}
}

\caption{\label{fig:pi-gtsp-grid} Global skew, local skew, rate multipliers and maximum local skew per node 
on the {\bf grid topology} for \textbf{AvgPISync} (left column) and \textbf{GTSP} (right column), respectively.}
\end{figure*}

Since fully distributed protocols exploit time information of the neighboring nodes to establish synchronization, it is fair to evaluate their performances on the grid topology. Figure \ref{fig:pi-gtsp-grid} shows global and local skews, rate multipliers and the maximum local skew observed per node for fully distributed AvgPISync and GTSP protocols on the grid topology, respectively. It can be observed that AvgPISync outperformed GTSP in terms of steady state local and global skews. On the other hand, the convergence time for GTSP, which employs distributed averaging for clock speed and offset adjustment, is faster than that of AvgPISync. This is due to the fact that, in GTSP, nodes are able to estimate instantaneous logical clock values of their neighboring nodes since they keep track of them. Using these estimates, they can calculate the average of the logical clock values of their neighboring nodes more precisely and hence they can catch-up the average neighborhood clock faster. However, since both protocols establish synchronization in a peer-to-peer, we conclude based on a few preliminary experiments that the convergence time of the rate multipliers and the maximum local skew per node will decrease as the density of the network increases. The results of the whole experiments are summarized in Table \ref{table:pi-gtsp}. Although the local skews of these protocols are quite comparable on this small grid topology, our simulations presented in subsection \ref{sec:simulations} shows that AvgPISync outperforms GTSP on larger networks.

\begin{table}

\center

\caption{ Summary of the clock skews in microseconds and convergence time in seconds with AvgPISync and GTSP on the grid topology. \label{table:pi-gtsp}}

\begin{tabular}{lcc}
 & \textbf{AvgPISync} & \textbf{GTSP} \\
\textbf{Max. Global}       	& {\bf 13}  & { 16} \\ 
\textbf{Max. Avg. Global }  & { \bf 9}  & { 12} \\ 
\textbf{Max. Local }        & { \bf 10}  & { 10} \\ 
\textbf{Max. Avg. Local }   & {\bf 4}   & { 5} \\ 
\textbf{Convergence Time }  & {\bf $\simeq$ 2000 }  & { $\simeq$ 1000}\\ 
\end{tabular}

\end{table}%

\subsection{Performance comparisons: FloodPISync and FTSP}

\begin{figure*}
\center
\subfigure[Global Skew]{
\includegraphics[scale=0.5]{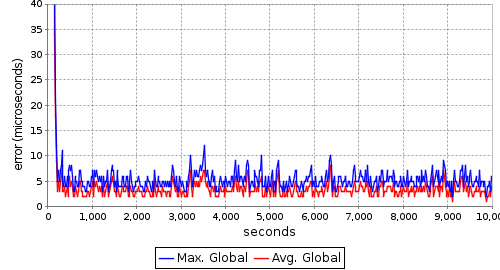}$\,
$\includegraphics[scale=0.5]{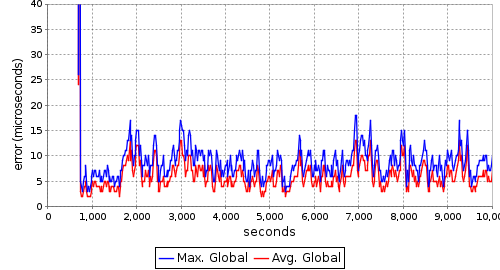}
}
\subfigure[Local Skew]{
\includegraphics[scale=0.5]{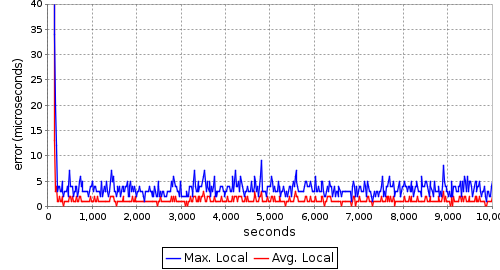}$\,
$\includegraphics[scale=0.5]{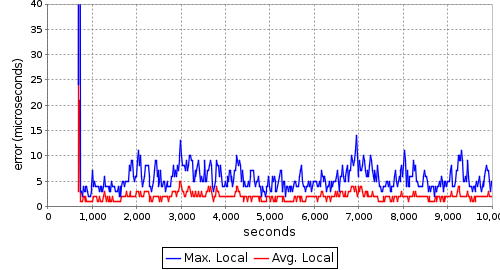}
}
\subfigure[Rate multipliers]{
\includegraphics[scale=0.5]{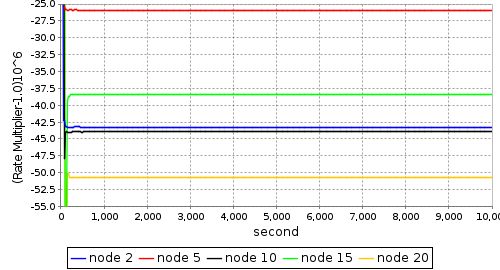}$\,
$\includegraphics[scale=0.5]{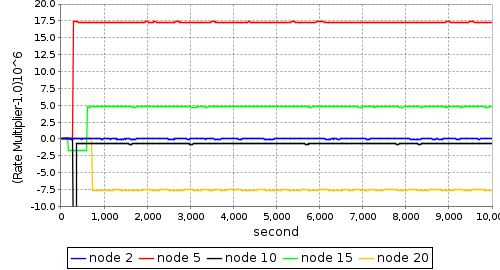}
}

\subfigure[Maximum skew to the reference node ]{
\includegraphics[scale=0.5]{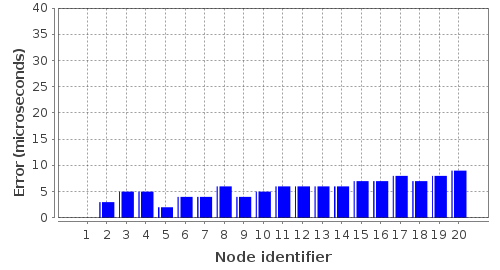}$\,
$\includegraphics[scale=0.5]{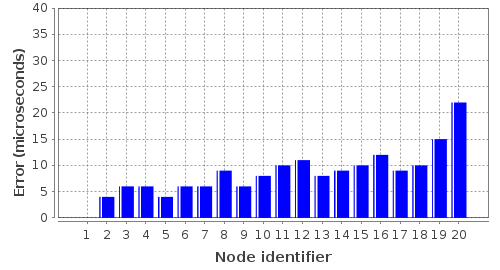}
}

\caption{\label{fig:pi-ftsp-grid} Global skew, local skew, rate multipliers and maximum skew to the reference node on the \textbf{grid topology} for \textbf{FloodPISync} (left column) and \textbf{FTSP} (right column), respectively.}
\end{figure*}

\begin{figure*}
\center

\subfigure[Global Skew]{
\includegraphics[scale=0.5]{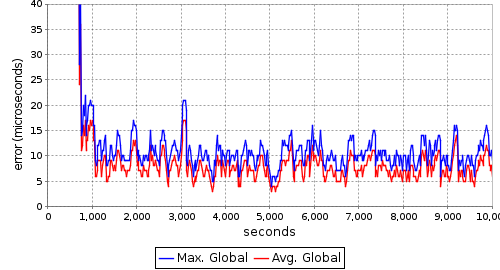}$\,
$\includegraphics[scale=0.5]{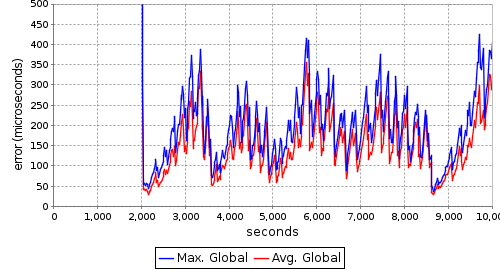}
}
\subfigure[Local Skew]{
\includegraphics[scale=0.5]{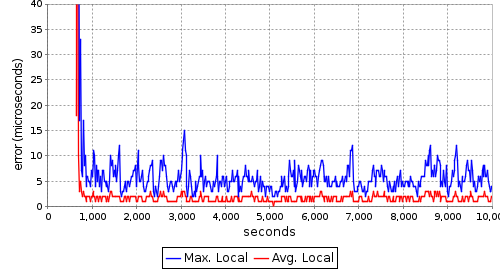}$\,
$\includegraphics[scale=0.5]{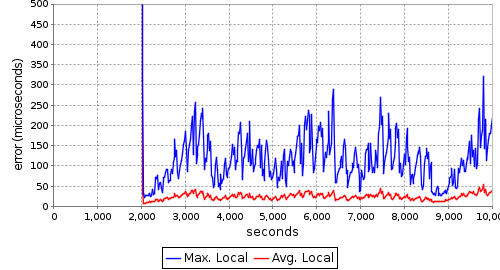}
}
\subfigure[Rate multipliers]{
\includegraphics[scale=0.5]{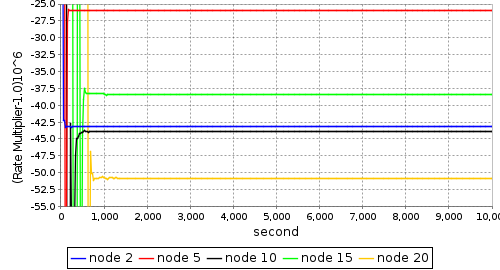}$\,
$\includegraphics[scale=0.5]{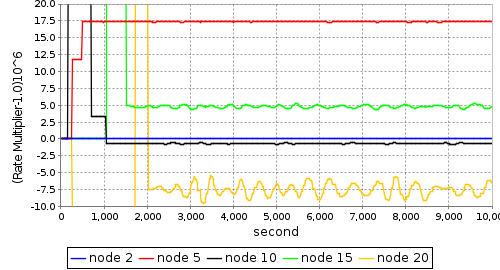}
}

\subfigure[Maximum skew to the reference node ]{
\includegraphics[scale=0.5]{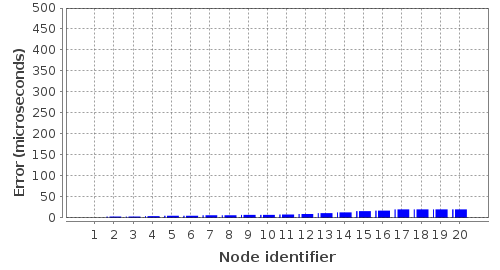}$\,
$\includegraphics[scale=0.5]{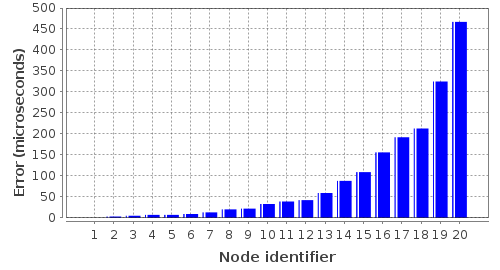}
}

\caption{\label{fig:pi-ftsp-line} Global skew, local skew, rate multipliers and maximum skew to the reference node on {\bf line topology} for \textbf{FloodPISync} (left column) and \textbf{FTSP} (right column), respectively.}
\end{figure*}

Figures \ref{fig:pi-ftsp-grid} and \ref{fig:pi-ftsp-line} show the skew values, rate multipliers and the maximum skew observed from the reference for FloodPISync and FTSP on the on the $5\times 4$ grid  and line topologies, respectively. It can be observed that in FTSP as the distance from the reference node 1 gets larger, the rate multipliers exhibit an oscillatory behaviour and this instability is reflected in the larger synchronization errors from the reference node. Due to the slow propagation of time information, waiting times at each hop increase the error of the reference time estimation. Even in a line topology of only 20 sensor nodes, we observed more than 0.5 milliseconds maximum global and local skew values with FTSP and an exponential growth of the synchronization error at each hop, which is consistent with the results presented in \cite{Lenzen2009Optimal}.

Differently, despite the use of the identical message pattern, FloodPISync protocol outperformed FTSP drastically in terms of clock skew. Especially on the line topology, it can be observed that the synchronization error grows very slowly unlike the exponential growth of FTSP, thus confirming the theoretical analysis in Section~\ref{sec:noise}. The rate multipliers in FloodPISync are quite stable and they do not exhibit fluctuations as the distance from the reference node increases. On the grid topology, there are many alternative paths for the time information of the reference node to be received by any sensor node. Even in this case, FTSP exhibited quite poor performance in terms of synchronization error. and it is outperformed by the FloodPISync protocol.

\begin{figure*}

\center

\subfigure[Global Skew]{
\includegraphics[scale=0.5]{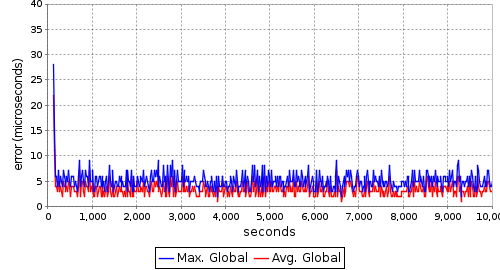}$\,
$\includegraphics[scale=0.5]{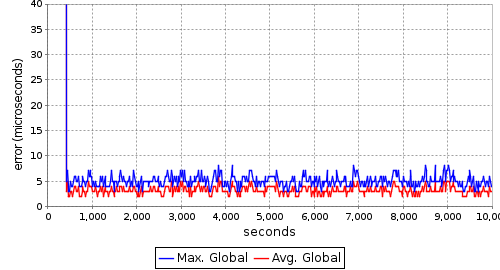}
}

\subfigure[Local Skew]{
\includegraphics[scale=0.5]{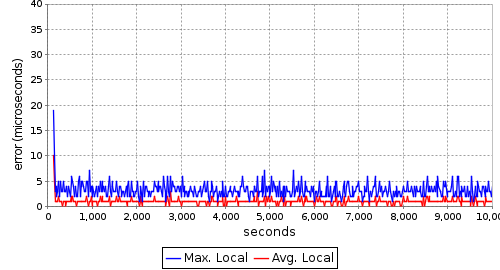}$\,
$\includegraphics[scale=0.5]{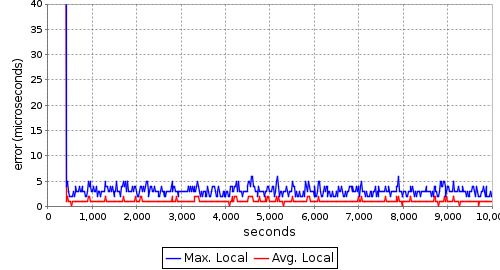}
}

\subfigure[Rate multipliers]{
\includegraphics[scale=0.5]{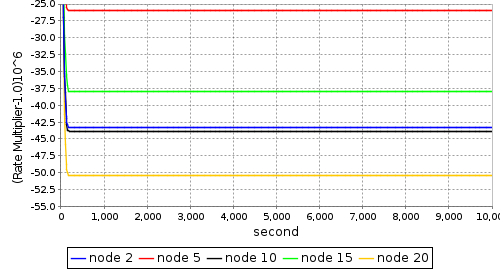}$\,
$\includegraphics[scale=0.5]{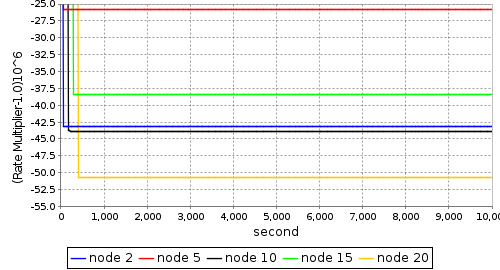}
}

\subfigure[Maximum skew to the reference node ]{
\includegraphics[scale=0.5]{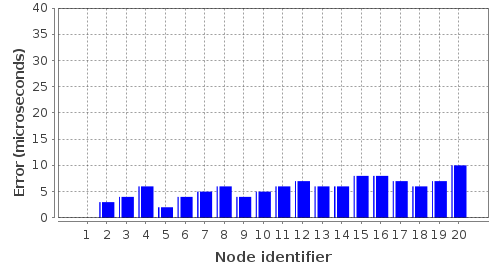}$\,
$\includegraphics[scale=0.5]{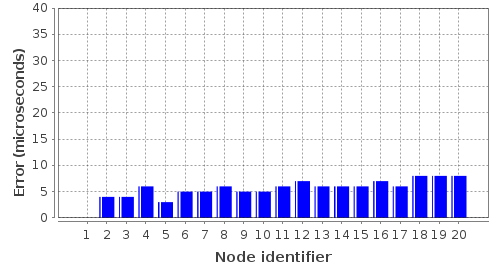}
}

\caption{\label{fig:pi-pulseSync-grid} Global skew, local skew, rate multipliers and maximum skew to the reference node on the {\bf grid topology} for \textbf{PulsePISync} (left column) and \textbf{PulseSync} (right column), respectively.}
\end{figure*}

\begin{figure*}

\center

\subfigure[Global Skew]{
\includegraphics[scale=0.5]{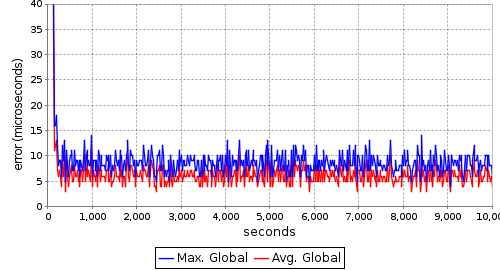}$\,
$\includegraphics[scale=0.5]{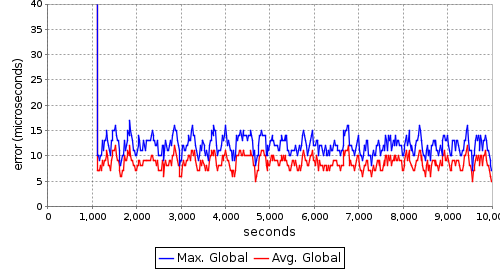}
}

\subfigure[Local Skew]{
\includegraphics[scale=0.5]{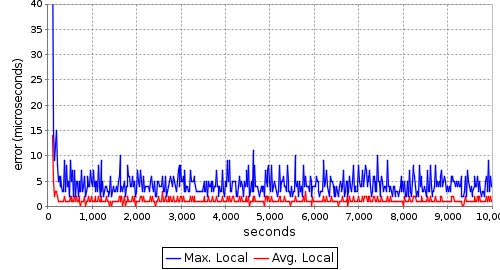}$\,
$\includegraphics[scale=0.5]{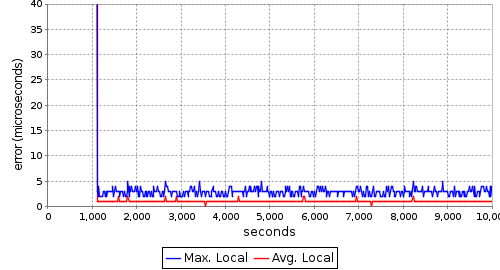}
}

\subfigure[Rate multipliers]{
\includegraphics[scale=0.5]{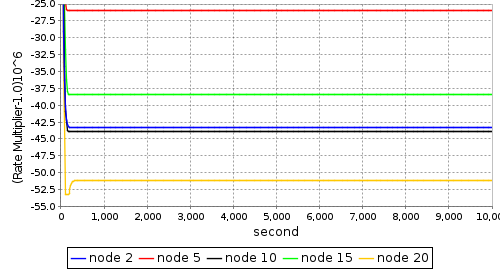}$\,
$\includegraphics[scale=0.5]{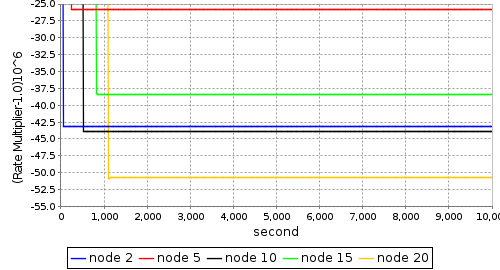}
}

\subfigure[Maximum skew to the reference node ]{
\includegraphics[scale=0.5]{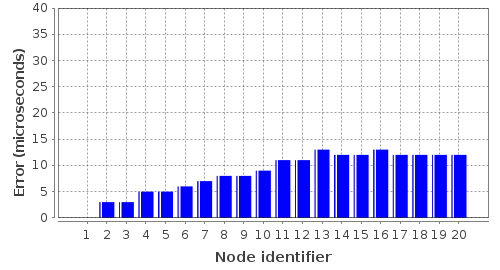}$\,
$\includegraphics[scale=0.5]{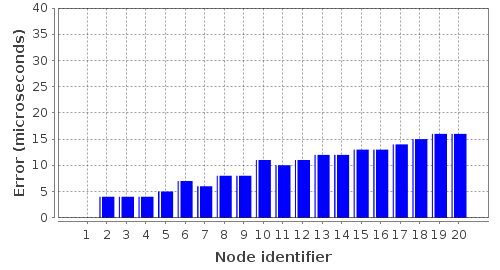}
}

\caption{\label{fig:pi-pulseSync-line} Global skew, local skew, rate multipliers and maximum skew to the reference node on the {\bf line topology} for \textbf{PulsePISync} (left column) and \textbf{PulseSync} (right column), respectively.}
\end{figure*}

\subsection{Performance comparisons: PulsePISync vs PulseSync}

PulseSync and PulsePISync propose reliable and fast flooding of the time information in order to reduce the effects of the waiting times. Figures \ref{fig:pi-pulseSync-grid} and \ref{fig:pi-pulseSync-line} show global and local skews, the rate multipliers and the maximum skew observed from the reference node measured for fast flooding based PulsePISync and PulseSync on the $5\times 4$ grid and line topologies, respectively. As we have shown in Section \ref{subsec:control_vs_least_squares} that the effect of various error sources appear in the time synchronization error dynamics as multiplicative noise with least-squares. Hence, it is desirable to propagate fresh time information as fast as possible through the pulses.  It can be observed that PulseSync reduced the fluctuations of the rate multipliers considerably when compared to FTSP since it employs a rapid flooding approach and reduces the amplification of the estimation errors at each hop.  With the same flooding speed, the performances of PulseSync and PulsePISync are quite comparable after steady state has reached. PulseSync exhibited a slightly better performance considering the average global and local skews. This is due to the fact that in least-squares regression a history is kept to calculate the estimated regression line. Thanks to this history, nodes are less sensitive to the small errors. However, due to the lack of any history in PulsePISync, nodes are more affected by these errors. The results of the experiments with flooding based protocols are summarized in Table \ref{table:ftsp-pulsesync-pi}.

\begin{table}

\center

\caption{Summary of the clock skews in microseconds and convergence time in seconds with FloodPISync, PulsePISync, FTSP and PulseSync on the grid and line topologies. \label{table:ftsp-pulsesync-pi}}

\begin{tabular}{lcccc||cccc}
\multicolumn{1}{l}{} & \multicolumn{2}{c}{\textbf{FTSP}} & \multicolumn{2}{c}{\textbf{FloodPISync}} & \multicolumn{2}{c}{\textbf{PulseSync}} & \multicolumn{2}{c}{\textbf{PulsePISync}} \tabularnewline
\multicolumn{1}{l}{}      & Grid & Line & Grid & Line & Grid & Line & Grid & Line \\
\textbf{Max. Global}       &  23  &  518  &  \bf 12  & \bf 21  & { 9} & { 17} & { \bf 10} & { \bf 14} \\ 
\textbf{Max. Avg. Global } &  20   & 476  & \bf 8   & \bf 17  & { 6} & { 12} & { \bf 8} & { \bf 10} \\ 
\textbf{Max. Local }       &  16   & 387  & \bf 9  & \bf 15   & { 6} & { 5}  & { \bf 8} & { \bf 12} \\ 
\textbf{Max. Avg. Local }  &  5   & 55   &  \bf 3   & \bf 4   & { 2}  & { 2}  & { \bf  3} & { \bf 3} \\ 
\textbf{Convergence Time}  & $  \simeq$ 750  &  $\simeq$ 2000 & { \bf < 500} & { \bf < 750} & { < 500}  & { $\simeq$ 1250}  & { \bf < 500} & { \bf < 500} \\ 
\end{tabular}
\end{table}%

As a summary of our overall experiments, it can be concluded that PISync protocols outperforms FTSP and GTSP, and has performance in terms of synchronization error which is comparable to PulseSync but with a significantly lower CPU and memory overhead. The fully distributed AvgPISync exhibits smaller local skew since each node synchronizes to their neighbouring nodes. For applications and protocols which require tight synchronization among the neighbouring nodes, this property is desirable. For external synchronization, flooding based PISync protocols are quite satisfactory. However, the best synchronization performance is achieved by employing fast flooding strategy with PulsePISync. For architectures which allows fast flooding, such as Glossy \cite{ferrari2011efficient}, PulsePISync is a good solution. 

\subsection{Reaction of PISync to dynamic changes in the network}

 Within the context of reaction to the network dynamics, the following main points should be considered: robustness against faulty (or bad) nodes broadcasting incorrect time information, and the perturbation of the network due to new nodes joining the network and due to instability of the clock frequencies caused by environmental factors such as instantaneous temperature changes.  

In our implementation of PISync, we followed a simple strategy to detect bad and faulty nodes, which is also implemented in the publicly available implementation of FTSP: When a node receives time information and observes that its synchronization error is greater than a predefined value\footnote{e.g. \textit{ENTRY THROWOUT LIMIT} in the FTSP implementation coming with TinyOS 2.1.2.}, it considers the received information as incorrect and discards it. However, if it successively observes this situation\footnote{i.e. if it observes 3 times successively in the FTSP implementation coming with TinyOS 2.1.2.}, this indicates that the reference time has changed and it decides to adapt its logical clock  to the received information.

\begin{figure}
\center

\includegraphics[scale=0.5]{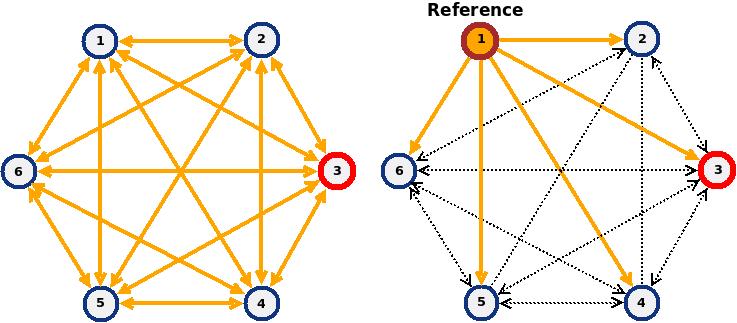}

\caption{\label{fig:pi-dynamics1} The fully connected topology of 6 Tmote Sky sensor nodes used for the observation of the reaction of the fully distributed AvgPISync (left) and flooding based FloodPISync (right) to the network dynamics. }

\end{figure}

In order to observe the reaction of the fully distributed and flooding based versions of the PISync to the network dynamics, we also performed experiments with AvgPISync and FloodPISync by turning some nodes off and then on at specific time instants. We used a fully connected topology of 6 sensor nodes as shown in Figure \ref{fig:pi-dynamics1}. For the FloodPISync, the node with identifier 1 is used as the reference node.

\begin{figure}
\center

\includegraphics[scale=0.5]{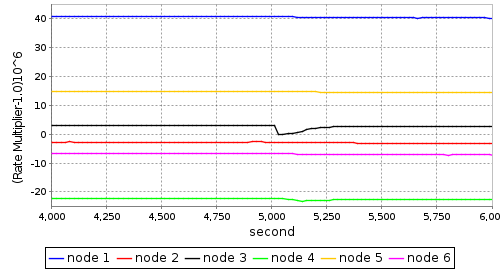}
\includegraphics[scale=0.5]{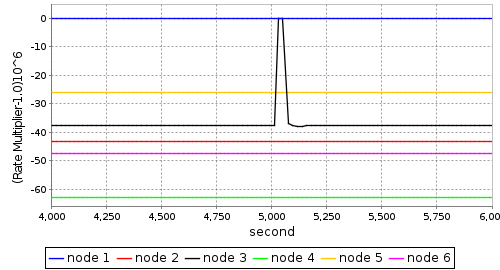}

\caption{\label{fig:pi-dynamics} Reaction of the fully distributed AvgPISync (left) and flooding based FloodPISync (right) to the network dynamics on the fully connected topology of 6 MICAz sensor nodes. }

\end{figure}

During the experiments with the AvgPISync, node 3 was turned off and then on after 5000 seconds from the beginning of the experiment, as shown in Figure~\ref{fig:pi-dynamics}. When this node started operating again, it listened its neighbours and tried to synchronize during a predefined fixed period of time. After this phase, it joined the network and started broadcasting its logical clock value periodically. As it can be observed, node 3 did not perturbed the logical clocks of its neighbouring nodes too much since they were already synchronized and communicating with each other. Moreover, since nodes do not keep track of their neighbouring nodes, no additional step was required to join the network. 

For the experiments with FloodPISync, we turned off and then on node 3 after 5000 seconds from the beginning of the experiment. Similarly, when the node 3 joins the network, it first try to achieve an initial synchronization by only collecting up-to-date time information of the reference node and it does not broadcast its logical clock for a predefined amount of time. After initial synchronization is achieved, it starts to broadcast its logical clock. Since nodes collect only the first synchronization message belonging to the current flooding round which is broadcasted by the reference node and discard the other messages, the newly joined node 3 did not disturb the network when it rejoined the network, as can be observed in Figure \ref{fig:pi-dynamics}. It should be noted that even in this small topology, a node may not collect time information from the reference node directly due to packet losses. Hence, there is a probability that any node may receive current time information of the reference node through a path of length two or more. If this path includes the newly joined node, network synchronization could be disturbed.

When FloodPISync and AvgPISync are compared in terms of the time took for the network to get adapted to the newly joined node 3, it can be observed that the re-synchronization period of the network with FloodPISync took approximately 150 seconds while it took approximately 250 seconds for AvgPISync. The adaptation time of AvgPISync is longer than that of FloodPISync since each sensor node tries to synchronize all of its neighboring nodes, thus increasing the adaptation time. On the contrary, in FloodPISync, each sensor node only considers the time information of the reference node.

The main factors effecting the frequency stability of the crystal oscillators are the environmental factors such as instantaneous temperature changes. For FloodPISync as well as for FTSP, PulseSync and many other flooding-based protocols in the literature, all nodes are treated equal and there is no distinction between the stable and unstable nodes. This issue has been discussed in \cite{Schmid:2010} for flooding based protocols. The authors proposed two solutions based on detecting stable neighbors and using time information flowing over them. These solutions can also easily be incorporated to FloodPISync to make it robust against unstable nodes. However, in order to select the most stable node among the neighbors, keeping track of the neighboring nodes and extra calculations are required.  For fully distributed protocols like GTSP and AvgPISync,  the issue of detecting unstable neighbors is overlooked in the literature. Discarding time information from unstable neighbors may be a simple solution. However, the network connectivity should be preserved in order to achieve network-wide synchronization. Hence, more sophisticated solutions are required that deserve to be a separate study itself. From our point of view, the solutions for the detection of the unstable nodes and adapting time synchronization should be generic and not specific to the time synchronization protocol.

In conclusion, PISync protocols are robust to changes in the network topology and re-synchronization is achieved quite rapidly without perturbing the logical clocks of the synchronized part of the network. However, FloodPISync  seems to be faster compared to the fully-distributed AvgPISync, thus making it more desirable in networks with frequent topological changes and node failures.

\subsection{Performance scalability in terms of network size}\label{sec:simulations}

The experimental results collected from small networks of 20 sensor nodes showed that FloodPISync outperforms FTSP drastically, PulsePISync catches the performance of PulseSync and AvgPISync outperforms GTSP in terms of synchronization error with significantly less computation and memory overhead. In order to observe and compare the performances of the aforementioned protocols for larger networks, we performed simulations for networks with different diameters. The error of time synchronization depends mainly on the network diameter as shown in \cite{Lenzen2009Optimal} and in Section~\ref{sec:noise}, therefore we employed simulations on grid   topologies having diameters ranging from 4 to 128 to observe the scalability of the protocols. 

\begin{figure}
\center

\includegraphics[scale=0.5]{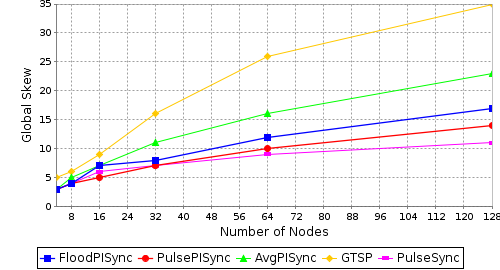}
\includegraphics[scale=0.5]{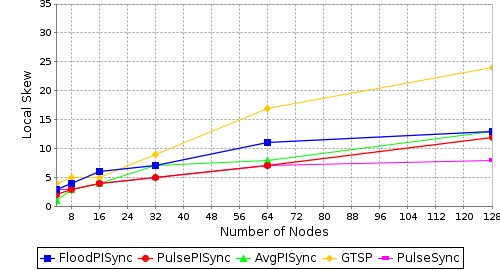}

\caption{\label{fig:simulations} Global skew (left) and local skew (right) values for PulseSync, GTSP, AvgPISync, FloodPISync and PulsePISync during simulations on networks with different diameters.}
\end{figure}

For simulations, we used JProwler \cite{JProwler} which is a discrete event simulator implemented in Java for prototyping, verifying and analyzing communication protocols of TinyOS ad-hoc wireless networks. JProwler models the important aspects of all levels of the communication channel. We used Gaussian wireless channel model of JProwler to simulate radio propagation. For MAC layer, we used the MAC protocol in JProwler, which is a variant of CSMA protocol implemented for Mica2 nodes. Since hardware clocks are not modelled in JProwler, we made some modifications and implemented constant drift hardware clock model, which models the drifts as uniformly distributed within the interval of $\pm$ 50 ppm. We implemented PulseSync, GTSP and PISync protocols in Java language to work with our modified version of JProwler. We performed 10 simulation runs for each diameter, which simulated a real-world experiment of $5\cdot10^5$ seconds (about 5 days). For each diameter, we averaged the maximum global and local skews of these 10 simulations, and the corresponding results are presented in Figure \ref{fig:simulations}.

Our simulation results are consistent with the results collected from our small testbed. Among PISync protocols, the local skew of the fully distributed AvgPISync protocol is smaller than that of the flooding based protocol FloodPISync on longer networks. This is evident since AvgPISync exploits all neighbourhood information and it strives to locally optimize the synchronization error. On the other hand, the global skew of AvgPISync is bigger than that of FloodPISync and PulsePISync since local optimization comes at the price of slow dissemination of time information. Flooding based protocols disseminate current time information faster. It can be concluded that AvgPISync is suitable for WSN applications which demand tightly synchronized neighbouring nodes, even in larger networks. FloodPISync and PulsePISync is more suitable for applications with the requirement of tight global skew. 

It can also be observed that AvgPISync performs better than GTSP in terms of both global and local skew on larger networks. For flooding based protocols, PulsePISync performs better than FloodPISync since it disseminates time information via fast flooding. During the dissemination of the time information in flooding based PISync protocols, the clocks of the nodes are instantaneously updated to the received time information\footnote{Line 6 of the Algorithm \ref{alg:PI} and Algorithm \ref{alg:PulsePI}.}. Hence, the logical clocks in PulsePISync are more sensitive to the small errors occurred due to message delays when compared to those in PulseSync, since PulseSync maintains a history, i.e. a regression table. Although the performance of PulseSync is slightly better than PulsePISync in larger networks due to this reason, it has considerable memory and CPU overhead as we mentioned previously.

In conclusion, the global and local skews of the PISync protocols grow very slowly with the network diameter. Hence, these protocols are quite scalable, and their synchronization performances are quite satisfactory in larger networks. 

\section{Conclusion and Future Work}

\label{sec:Conclusion-and-Future}

In this article, we took into consideration recent flooding based and fully distributed time synchronization protocols in the WSN literature. We emphasized that these protocols are heavy in terms of computation and memory allocation overhead, need to keep track of time information of the other nodes which is a serious shortcoming for dense networks, require large amount of information to be exchanged among the sensor nodes, and have poor performance scalability. We proposed a new control theoretic distributed time synchronization algorithm, named PISync, which is based on a Proportional-Integral (PI) controller. We presented flooding-based and fully distributed PISync protocols which are based on PISync algorithm and observed their performances through real-world experiments and simulations. We revealed that PISync protocols have several superiorities over existing protocols: (i) they do not store any distinct time information and have very little memory allocation overhead, (ii) they have very little CPU overhead, (iii) they require very little amount of information to be exchanged, (iv) they have very small code footprint, (v) they are  quite scalable in terms of steady state global synchronization error performance. Future work may include employing PISync algorithm for TDMA applications in order to control sleep and wake-up schedules of the sensor nodes to achieve very low-power consumption. Observing the behaviour of PISync in mobile networks is also interesting future research avenue.

\section*{{\large Acknowledgments}}
This work is supported by the European Community's Seventh Framework Programme [FP7/2007-2013] under grant agreement n. 257462 HYCON2 Network of excellence.

\bibliography{references}

\end{document}